\pdfoutput=1
\documentclass[11pt,a4paper]{article}
\usepackage{jheppub}

\usepackage{enumerate}	
\usepackage{hyperref}	
\usepackage{amsmath}
\usepackage{slashed}
\usepackage{amssymb}
\usepackage{euscript}
\usepackage{bbm}
\usepackage{bm}
\usepackage{graphicx}
\usepackage{subfigure}%for subnumbering inside a figure

\def\alf{\alpha_}

\newcommand{\bk}{{\bf{k}}}
\newcommand{\ba}{{\bf{a}}}
\newcommand{\Li}{\text{Li}}

\newcommand{\Pbbm}{\mathbbm{P}}
\newcommand{\Ibbm}{\mathbbm{1}}

\newcommand{\Peu}{\EuScript{P}}

\newcommand{\veps}{\varepsilon}

\subheader{\bf IFJPAN-IV-2009-6\\
    CERN-PH-TH/2011-032}
\title{Two real parton contributions to non-singlet
       kernels for exclusive QCD DGLAP evolution}
\author[a,b]{S. Jadach,}
\author[a]{A. Kusina,}
\author[a]{M. Skrzypek,}
\author[a]{M. Slawinska}
\affiliation[a]{Institute of Nuclear Physics, Polish Academy of Sciences,\\
  ul.\ Radzikowskiego 152, 31-342 Cracow, Poland.}
\affiliation[b]{CERN, PH Division, TH Unit, CH-1211 Geneva 23, Switzerland.}
\abstract{Results for the two real parton differential distributions
needed for implementing a next-to-leading order (NLO) parton shower
Monte Carlo are presented.
They are also integrated over the phase space in order to
provide solid numerical control of the MC codes
and for the discussion of the differences between the standard
$\overline{MS}$ factorization and Monte Carlo implementation
at the level of inclusive NLO evolution kernels.
Presented results cover the class of non-singlet
diagrams entering into NLO kernels.
The classic work of Curci-Furmanski-Pertonzio
was used as a guide in the calculations.}
\arxivnumber{1102.5083}

\begin{document}
\maketitle

%%%%%%%%%%%%%%%%%%%%%%%%%%%%%5
%%%%%%%%%%%%%%%%%%%%%%%%%%%%%5
\section{Introduction}
%%%%%%%%%%%%%%%%%%%%%%%%%%%%%5
%%%%%%%%%%%%%%%%%%%%%%%%%%%%%5

With the start of the operation of the Large Hadron Collider (LHC) in CERN
and the progress in the analysis of growing data samples for many
scattering processes, mastering the
precise evaluation of strong interactions effects
within perturbative Quantum Chromodynamics
(QCD)~\cite{Gross:1973ju,Gross:1974cs,Georgi:1951sr}
will become quickly more and more important.
QCD is a well established theory --
testing the validity of QCD is not an open issue any more.
Its principal role in the LHC data analysis will be providing
precise predictions for rates and distributions of quarks,
gluons and hopefully other newly found particles carrying
colour charge,
being part of either signal process or background.

The most important theoretical tools in perturbative QCD (pQCD)
calculations, apart from Feynman diagrams, renormalization, etc.
are the so called factorization theorems,
see for instance~\cite{Ellis:1978ty,Collins:1984kg,Bodwin:1984hc},
which allow to describe any scattering process with
a single large mass or transverse momentum scale $\mu_F$
(enforcing short distance interaction),
in terms of the on-shell hard process
matrix element (ME) squared and convoluted with the {\em ladder parts}.
The hard process is calculated up to a fixed perturbative order.
The ladder parts are defined
for each colored energetic parton entering (exiting)
the hard process.
The initial state ladders are conveniently
encapsulated in the {\em inclusive} parton distribution functions, PDFs.

The logarithmic dependence of the parton distribution functions (PDFs)
on the large scale $\mu_F$
is described as a DGLAP~\cite{DGLAP} evolution of the PDFs.
This evolution was studied for the inclusive PDFs up to NLO level
in the early 80's, see refs.~\cite{Floratos:1978ny,Curci:1980uw},
and was recently established at the NNLO level~\cite{Moch:2004pa,Vogt:2004mw}.

Instead of being encapsulated in the inclusive PDF,
the multi-parton emission process of the initial state ladder,
can be modelled using direct stochastic
simulation in terms of four-momenta and other quantum numbers,
within the Monte Carlo (MC) parton shower.
Here, the baseline works have been done in mid-80's, 
see refs.~\cite{Sjostrand:1985xi,Webber:1984if},
where the leading order (LO)
ladder was implemented in parton shower MC (PSMC) programs.
Standard LO level PSMCs implement also the process
of hadronization of the light quarks and gluons
into hadrons and play an important role
in the software for all collider experiments because of that.

Fulfilling the challenging requirements on the quality and precision
of the pQCD calculations,
needed for the experimental data analysis at the LHC,
enforces for the first time an urgent solution of the problem
of upgrading exclusive PSMC to the complete NLO level, the same level,
which was reached for inclusive PDFs two decades ago.
This task is highly nontrivial
mainly because the classic factorization
theorems~\cite{Ellis:1978ty,Collins:1984kg,Bodwin:1984hc}
were never designed for the exclusive MC implementation,
but rather for defining inclusive PDFs
and performing fixed order calculations for the hard process,
convoluted with these PDFs.

Let us comment briefly on the longer term physics impact of the present work.
Remembering that QCD is not any more a new theory, the main
impact of this work will be  the improvement of calculations of QCD effects,
for hadron collider experiments like LHC,
with the aim of improving chances of discovering directly or
indirectly New Physics and/or better measurement
of the Electroweak Standard Model parameters,
especially when high statistics, high precision data are accumulated.

More precisely, this work elaborates on pQCD effects in the initial state,
which from the perspective of the LHC experiments influence mainly:
(A) overall normalization of the hard processes
through parton luminosities,
(B) distributions of transverse momenta ($k_T$) of incoming partons
deforming many other important distributions in any hard process,
including searches for supersymmetry, etc.,
(C) and provide one or more jets accompanying hard process.

The longstanding problem in the data analysis is that
the above three classes of important QCD effects
are addressed by three separate theoretical
pQCD calculational tools, based on different incompatible perturbative techniques:
(A) strictly collinear NLO PDFs~\cite{DGLAP},
(B) semi-inclusive schemes of infinite order
soft gluon $k_T$ resummation~\cite{Collins:1984kg,Kulesza:1999sg,Kulesza:2002rh}
or, alternatively, LO parton shower 
Monte Carlos~\cite{Sjostrand:1985xi,Marchesini:1988cf}
(C) finite order NLO (NNLO) 
calculations~\cite{Altarelli:1979ub,Anastasiou:2003ds}
sometimes combined with a LO parton shower 
MC~\cite{Frixione:2002ik} or collinear PDFs.

In the analysis of experimental data combining
these (and other) techniques is not only inconvenient,
but also is a serious source of irreducible
theoretical (and experimental) uncertainties.
This old and well known problem becomes more severe
with the increasing precision of the collider data,
and will inevitably plague high statistics, high precision LHC data.
The ultimate aim of the present work is to provide a basis
for designing a single Monte Carlo
program addressing all three classes of the above pQCD effects at once,
within the same consistent pQCD theoretical framework.
However, for this to be realized one has to start with solving
one basic difficult problem --
the upgrade of the initial state parton shower MC to at least the same
level as standard PDFs, that is to the NLO level,
in the fully exclusive manner.
The present work provides essential building blocks (exclusive kernels) for
this critical extension of the parton shower, and analyzes factorization
scheme differences with respect to the standard CFP $\overline{MS}$ scheme.
While this work focuses on the implementation of NLO
corrections to the initial state ladder parts (parton showers),
the hard process part at NLO within the same scheme
is discussed in ref.~\cite{Jadach:2011cr}%
\footnote{ A complementary approach can be found in
  refs.~\cite{Ward:2007xc,Joseph:2009rh,Joseph:2010cq}.}.

The critical problems to be solved on the way to NLO PSMC are the following:
\begin{enumerate}
%%%%%%%%
\item {\bf Violation of four momentum conservation.}
%%%%%%%%
In the standard collinear factorization four momentum conservation is broken both
between the hard process and the ladder as well as between
the ladder segments (2PI kernels).
%{\em Ad 1:}
The source of this non-conservation
in the standard collinear factorization is the introduction of
the {\em projection operators},
$P_{kin}$  in ref.~\cite{Ellis:1978ty}, 
operator $\Pbbm$ in ref.~\cite{Curci:1980uw},
or operator $Z$ in ref.~\cite{Collins:1998rz}.
These operators are absent
in the first step of the separation of the collinear
singularities into the ladder parts --
they are introduced later on in order to 
(i)  conveniently isolate the lightcone variable integration
out of the phase space for analytical integration and
(ii) facilitate the order-by-order pQCD calculations separately
for the hard process ME and the ladder elements (kernels).
This non-conservation happens in the transverse momenta, 
which are anyway treated inclusively (integrated over),
hence this bad feature of collinear factorization goes almost
unnoticed.%
\footnote{
 Except of ref.~\cite{Collins:1998rz},
 where it is discussed in a more detail.}
In the traditional LO PSMCs
the above non-conservation is repaired 
``by hand''~\cite{Sjostrand:1985xi,Webber:1984if},
but in such a way that the NLO effects induced by this
reparation are analytically almost uncontrollable.
This is not a problem, unless one attempts to complete NLO in the ladder,
or to combine an NLO hard process ME with
a LO PSMC~\cite{Frixione:2002ik}.
A systematic solution of the above problem must involve replacing
the projection operator $P_{kin}$ by a more sophisticated
operation involving a special parametrization
of the entire phase space for the hard process and the ladders.
An explicit example of such a solution for the $W/Z$ production process
(Drell-Yan process) can be found in ref.~\cite{Jadach:2011cr}.
%%%%%%
\item {\bf Separation of singularities before phase-space integrations.}
%%%%%%
In the collinear factorization, separation of the LO singular contributions
in the form of the leading logarithm $\ln\frac{Q^2}{m^2}$
or $\frac{1}{\veps}$ poles of dimensional regularization can be done
only
{\em after} the phase space integration. In order to construct an
efficient Monte Carlo algorithm this separation has to be done at the
very beginning, 
at the integrand level, {\em before} the phase space integration.
This requires going beyond the inherently {\em inclusive} approach of
collinear factorization.  
%{\em Ad 2:}
% The problem of the traditional collinear factorization
% being inherently inclusive,
% and therefore not suitable for the MC is widely known and partly solved.
The effort of getting the NLO prediction for the semi-inclusive distribution
in the phase space of the hard process
(like $k_T$ and rapidity distributions in $W/Z$ production)
started already quite early,
see for instance~\cite{Altarelli:1979ub,Altarelli:1984pt}.
More recently, Monte Carlo tools
combining NLO ME for the hard process with the LO PSMC
were developed in~\cite{Frixione:2002ik} and \cite{Nason:2004rx}.
Studies on redefining PDFs in a partly exclusive form beyond LO
(unintegrated PDFs)~\cite{Collins:1984kg},
or in exclusive form (fully unintegrated PDFs)~\cite {Collins:2007ph}
are also pursued.
%%%%%%
\item {\bf Negative ``probability distributions''.}
%%%%%%
The NLO corrections in collinear factorization are negative in some
regions of phase space
and therefore {\em cannot} be generated directly in the Monte Carlo,
if we insist on the realistic simulation using positive-weight MC events.%
\footnote{Unless one admits the painful scenario with negative weight MC events,
  as in ref.~\cite{Frixione:2002ik}.}
%{\em Ad 3:}
The reasons for non-positiveness of the NLO corrections is well understood.
For instance, in the physical gauge positive squares of the Feynman
diagrams are typically more divergent and form the LO approximation,
while non-positive interferences
are collected in NLO corrections.
The use of kinematic projection operators in factorization
and related subtractions are another sources.
The factorization procedure collects all these non-positive
corrections into separate objects and the
integration over the phase space is done for each of them separately.
%Naively one could attempt to create
%a separate branch of the MC algorithm 
%for non-positive NLO distributions
%-- this is not possible simply because they are not positive%
Consequently, part of the factorization procedure has to be reversed
in order to recombine non-positive NLO distributions with the
positive LO distributions, before the MC algorithm is designed.
The above {\em defactorization} procedure has to be outlined.
An example proposal relying
on Bose-Einstein symmetrization was formulated
in ref.~\cite{Jadach:2010ew}
and an even more promissing one is described in ref.~\cite{Jadach:2010aa}
(similar to that in ref.~\cite{yfs:1961}).
%%%%%%
\item {\bf Lack of the published exclusive NLO distributions.}
%%%%%%
We have not found exclusive NLO distributions
forming the NLO corrections in the ladder part in literature --
all published results
in the context of the NLO calculations of kernels for evolution of PDFs
are integrated over the phase space.
The main objective of this paper is to provide such distributions for
the non-singlet case.
%%%%%%
\item {\bf Inclusive treatment of multigluon soft limit.}
%%%%%%
%Inclusive treatment of multigluon soft limit in
%the existing NLO corrections to PDFs.
%{\em Ad 5:}
One of the critical issues in any type of collinear factorization
is the behavior of many-gluon distributions in the soft limit.
% In certain sense it is simpler in the fraamework of the inclusive
% collinear factorization and requires much more attention
% and effort in the exclusive factorization of the NLO PSMC.
In the inclusive approach it is enough to know that
the cancellations between real and virtual soft contributions
allow us in principle to neglect entire
classes of diagrams and/or divergent contributions
-- they sum up to zero.
In the exclusive MC approach these contributions/singularities,
instead of being dropped out,
have to be modelled precisely to infinite order.
The above soft gluon behavior is rather well known,
albeit more complicated in QCD than in QED due to the
(non-abelian) triple gluon vertex --
the eikonal limit and angular ordering 
are known to govern it~\cite{khoze-book}.
% In the present work soft limits of all NLO distributions
% will be worked out and analyzed carefully.
In ref.~\cite{Slawinska:2009gn} a detailed analysis of the
soft limit for the $C_F^2$ and $C_FC_A$ contributions
for non-singlet evolution kernels was performed
and the well known angular ordering was exposed,
both numerically and analytically, for the first time 
using exact double gluon distributions presented explicitly.
% In the present work this discussion will be slightly extended.
% and the distributions used in ref.~\cite{Slawinska:2009gn}
% will be given explicitly.
%%%%%%
\item {\bf Inappropriateness of $\overline{MS}$ factorization scheme.}
%%%%%%
The issue of factorization scheme dependence has to be revisited and
one has to decide about the best factorization scheme (FS) for
the PSMC implementation. It should not be taken for granted
that it will be the $\overline{MS}$ scheme, which
is presently the ``industry standard'' for the inclusive PDFs
%
%{\em Ad 6:}
%Exploration of the factorization scheme (FS) dependence in the
%traditional inclusive version in early 1980's has resulted
%in the universal adoption of 
and in pQCD calculations for the hard process.%
\footnote{Since then the FS dependence is reduced to
 the discussion of the residual dependence on the factorization scale 
 $\mu_F$ of $\overline{MS}$ scheme due to higher orders.}
%For NLO evolution of PDFs in the $\overline{MS}$ scheme 
%see ref.~\cite{Curci:1980uw}.
In the view of the above the following question emerges:
Can one stay strictly within the $\overline{MS}$ scheme
for the MC implementation of the NLO ladders combined
with the NLO hard process ME?
In the present work we will addres this question partly,
for the MC modelling of the ladder part.
It is studied for the hard process ME
in ref.~\cite{Jadach:2011cr}.
Let us indicate that our answer will be negative:
FS of the NLO MC {\em has to differ} from $\overline{MS}$
scheme for several reasons.
Some of them will be discussed in detail
in this paper.
% Here, we will pay attention to any relevant aspect of 
% the presented analytical results,
% clearly indentify and describe all elements needed for 
% the discussion of the above question, with the aim
% of precise determination of the differences
% between the standard $\overline{MS}$ scheme
% and the FS actually implemented in the NLO PSMC.
%%A few general remarks may clarify an emerging picture.
%Hard process ME part and the ladder parts are separately FS-dependent.
%Their phase space convolution is, however,  FS-independent.
%Once choice of FS is made for the hard process,
%then FS for the ladder part gets fixed and vice versa.
The key element defining FS are the so-called soft collinear counterterms.%
\footnote{We refer to the exclusive version of soft collinear counterterm, 
  which is the distribution within the 1-particle
  phase space encapsulating collinear (and soft) singularity --
  not its integral
  as in inclusive FS.}
In MC we choose them to be identical with the distribution used in the LO MC,
e.g. the LO level PSMC is constructed by means of iterating soft
  collinear counterterms.
This choice on one hand will simplify NLO corrections in the MC,
but it will also cause departure from the $\overline{MS}$ FS.
The second source of discrepancy will be the fact
that in the MC, the factorization scale will be identical to
a well defined kinematic variable $Q_F$ of the LO MC,
replacing the formal parameter $\mu_F$ of the $\overline{MS}$ FS.
The logarithm of $Q_F$ is then used in the MC as
 an ordered {\em evolution time} variable.
$Q_F$ will typically be maximum transverse momentum,
maximum angle, or virtuality of the emitted particles.
For the gluonstrahlung, 
the choice of the maximum transverse momentum
results in the MC FS very close to
$\overline{MS}$ FS.%
\footnote{This fact is well known,
 see also analysis of ref.~\cite{Kusina:2010gp}.}
However, the soft gluon eikonal limit for
contributions like gluon pair production or quark-gluon transitions,
dictates the use of the variable related to the maximum angle (rapidity)
as a factorization scale variable in the MC.
Third reason is the presence in the $\overline{MS}$ FS
of certain artifacts of dimensional regularization
which cannot be implemented in the MC in four dimensions.
%%%%%%
\item {\bf Constrained evolution.}
%Lack of new versatile Monte Carlo techniques implementing constrains
Constrains
on the momenta and other quantum numbers imposed by the hard process on
the initial state parton shower (ladder),
especially important in the presence of the narrow resonances, must be
taken into account by the PSMC.
The clever and powerfull MC technique of backward evolution of
ref.~\cite{Sjostrand:1985xi} is good for the LO PSMC,
but for the NLO case one may need something more sophisticated.
%\subsection{State of the art}
%%%%%%%%%%%%%%%%%%%%%%%%%%%%%%%%
%Going point by point in the above list we summarize on the present situation:
%{\em Ad 7:}
The dependence of the ladder part (PDF) on the factorization scale
is described in pQCD by the integro-differential 
DGLAP~\cite{DGLAP} evolution equation.
Its solution has the form of a time ordered (T.O.) exponential
with the ordering in the logarithm of the factorization scale.%
\footnote{Running of the coupling constant is included
  with the help of the usual renormalization group argument.}
This T.O. exponential can also be obtained directly from
the ladder Feynman diagrams.
In the Monte Carlo T.O. exponential is conveniently
modelled using a Markovian MC algorithm.%
\footnote{
 Known already since prehistory of the Monte Carlo methods
 at Los Alamos National Laboratory~\cite{Everett-1972}.
}
%In case of more a complicated evolution kernel, MMC is supplemented
%with the convenient veto method~\cite{Sjostrand:2006za,Jadach:2003bu} (???).
However, a Markovian MC algorithm would be highly inefficient
for modelling the initial state radiation (ISR) ladder,
when hard process selects very narrow range of energies
and flavors of the partons incoming into hard process
(with narrow W/Z resonances).
Here, the backward evolution MC algorithm is a standard
solution~\cite{Sjostrand:1985xi} --
used in all standard LO PSMCs.
It is to be seen whether backward evolution MC is upgradable to NLO PSMC.
In the meantime, the constrained MC
algorithm of refs.~\cite{Jadach:2005bf,Jadach:2005yq}
offers an interesting alternative.
While the backward evolution MC needs pretabulated solutions (PDFs) of the
evolution equations, which have to be prepared beforehand
using non-MC auxiliary codes,
the constrained MC performs evolution on its own.	
% Altogether, it looks that the toolbox of the MC methods and algorithms
% available for constructing NLO PSMC is quite versatile and complete.
\end{enumerate}

In this article we mainly addres point 4
in the above list of problems. However also point 6 is discussed
in some detail.

Let us now establish precise terminology
concerning diagrams and phase space integration.
We will elaborate on the diagrams contributing to
non-singlet evolution kernels at the NLO level.

Generally, in the calculation of the exclusive/inclusive
evolution kernels in this work we will take
paper of Curci Furmanski and Petronzio~\cite{Curci:1980uw} (CFP)
as the starting point and as the reference in all NLO calculations.
For the non-singlet part of the QCD DGLAP evolution
we will calculate (or recalculate) both the standard inclusive NLO kernels
and the new {\em exclusive} (unintegrated) ones,
which are needed for constructing NLO PSMC.
This work prepares building blocks
for NLO PSMC, whereas the actual MC algorithm and all issues
related to the factorization scheme used in NLO PSMC will
be discussed in a separate paper~\cite{Jadach:2011cr}.

\begin{figure}[h!]
\begin{centering}
\subfigure[]{
  \includegraphics[height = 2.5cm]{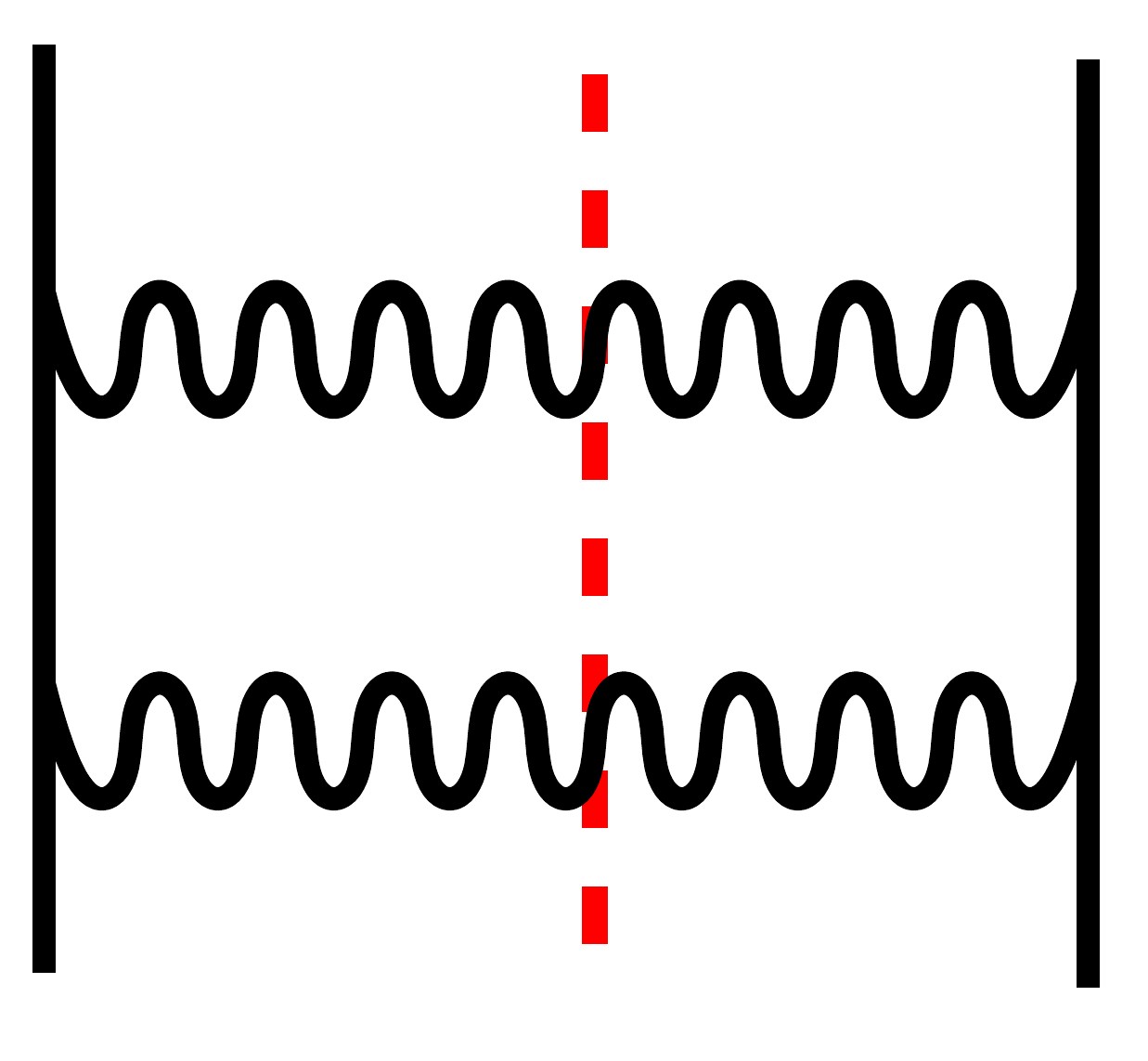}
\label{subfig:Br}
}
\subfigure[]{
  \includegraphics[height = 2.5cm]{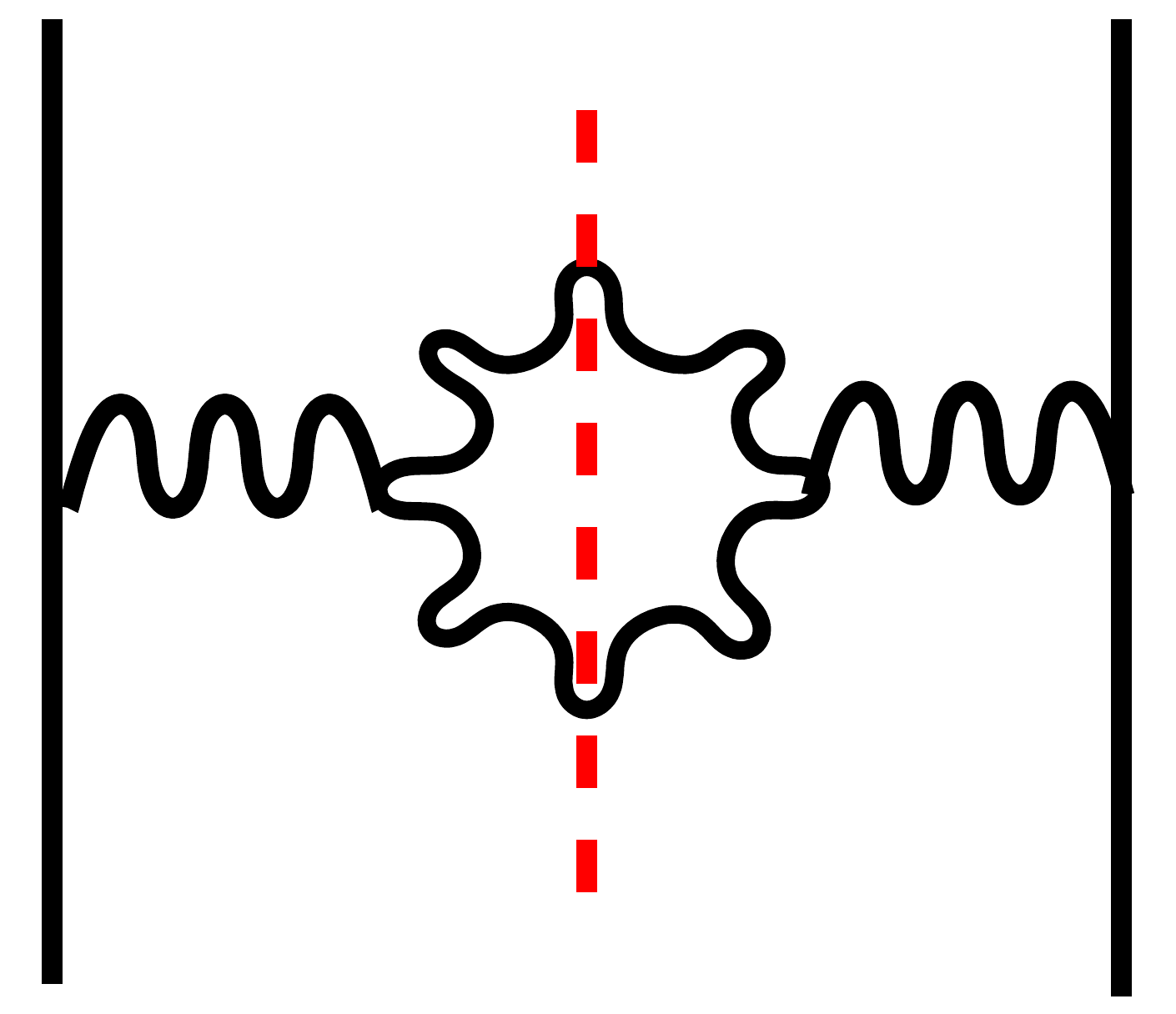}
  \label{subfig:Vg}
}
\subfigure[]{
  \includegraphics[height = 2.5cm]{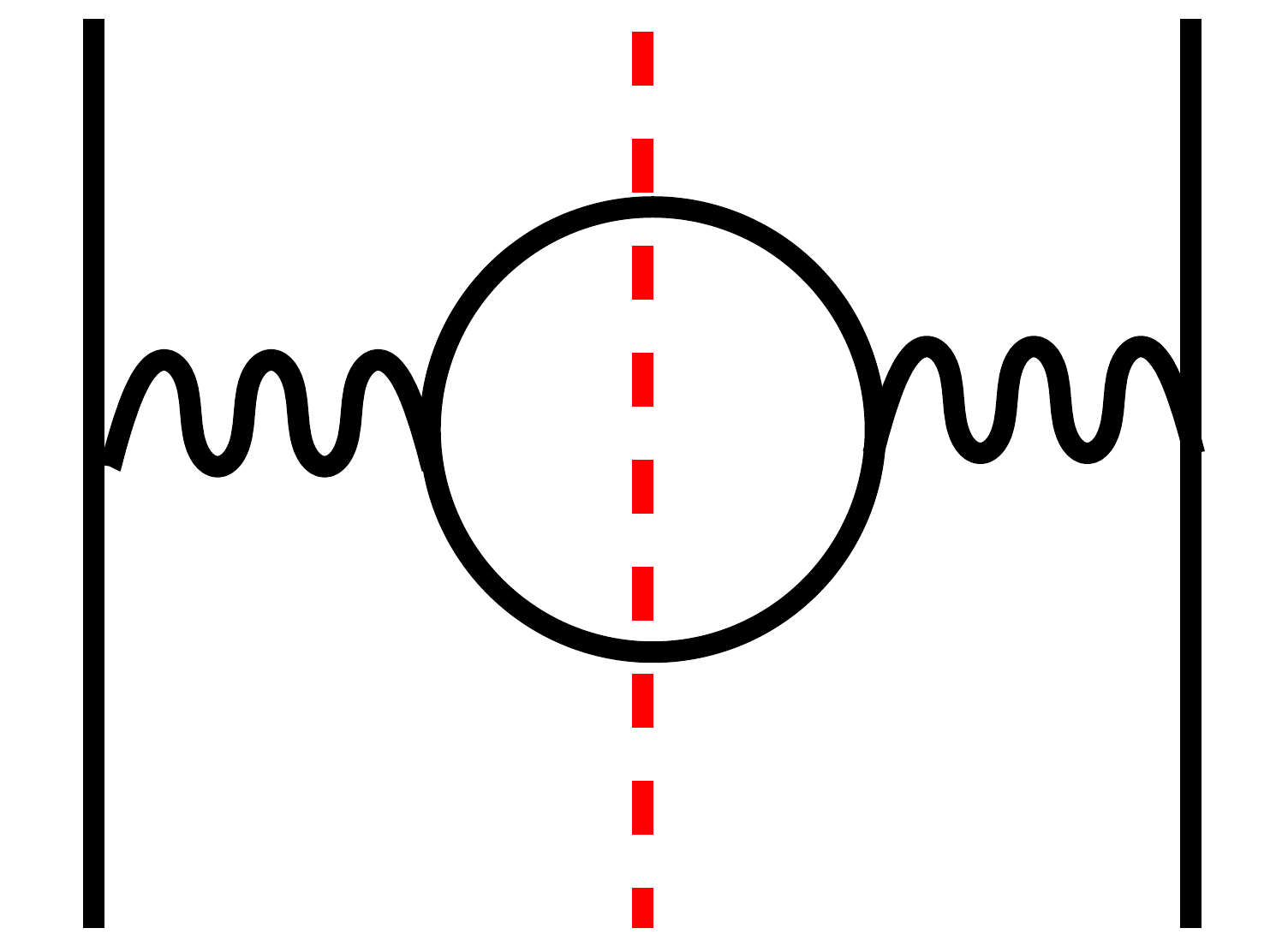}
  \label{subfig:Vf}
}\\
\subfigure[]{
  \includegraphics[height = 2.5cm]{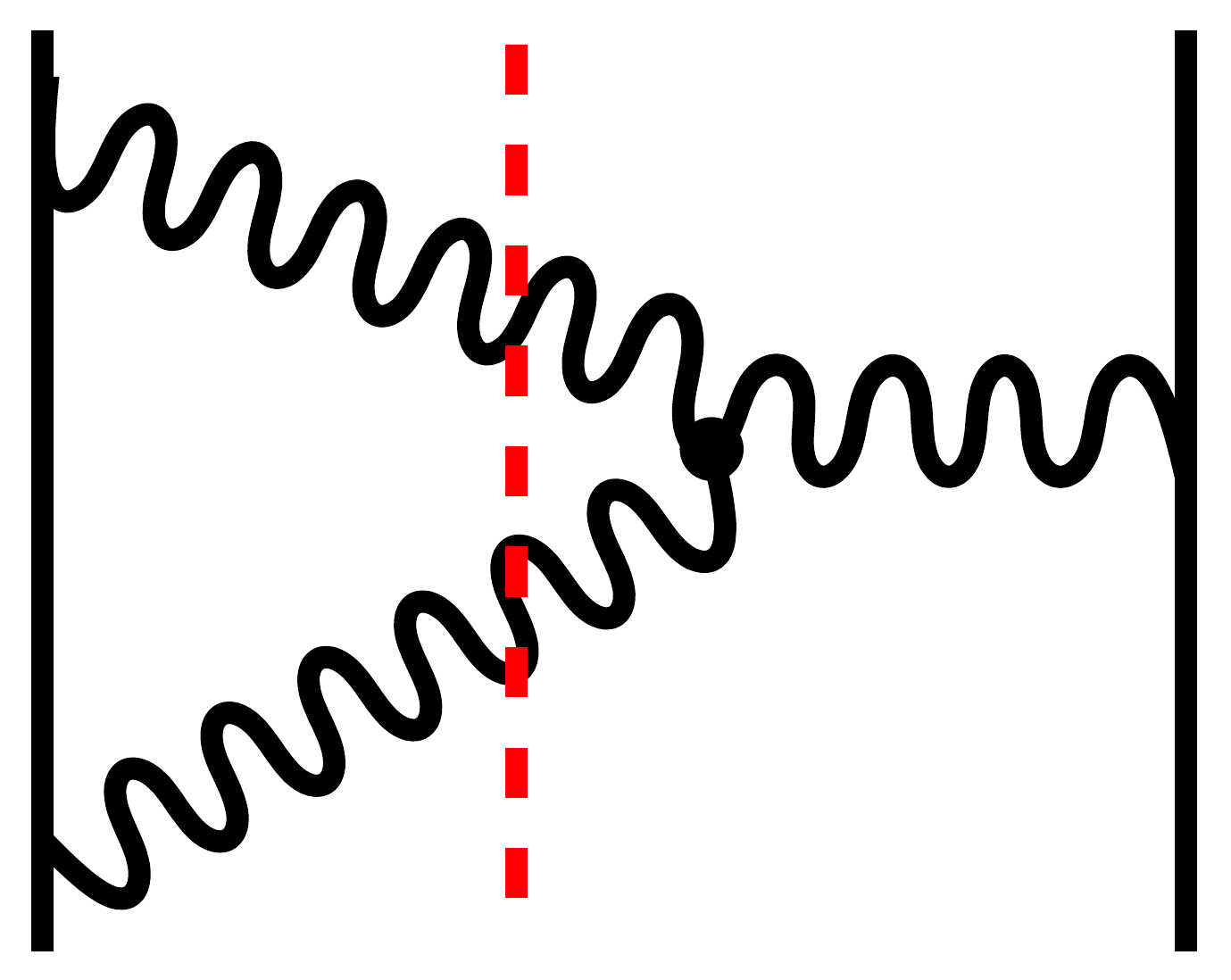}
  \label{subfig:Yg}
}
\subfigure[]{
  \includegraphics[height = 2.5cm]{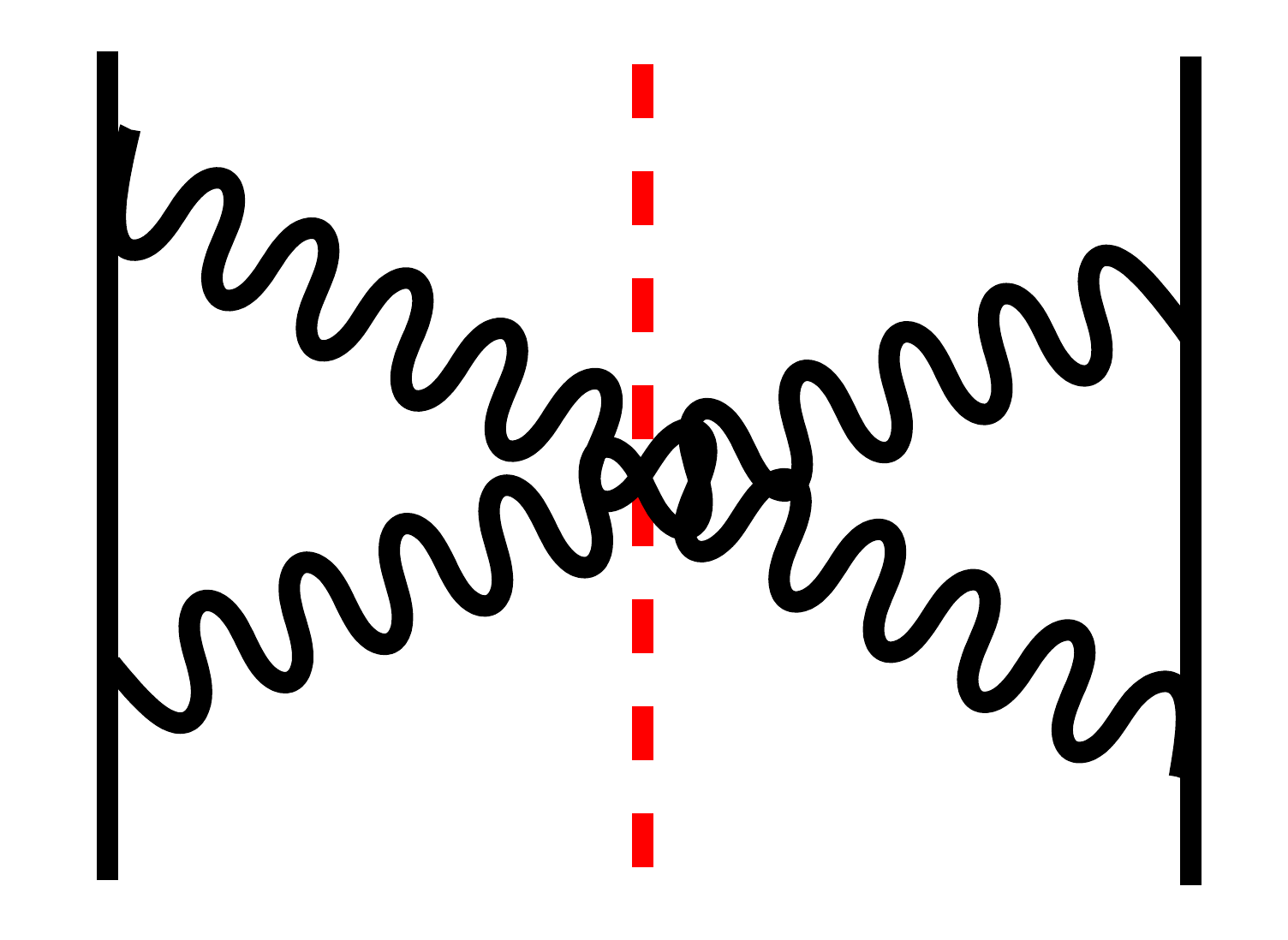}
  \label{subfig:Bx}
}
\subfigure[]{
  \includegraphics[height = 2.5cm]{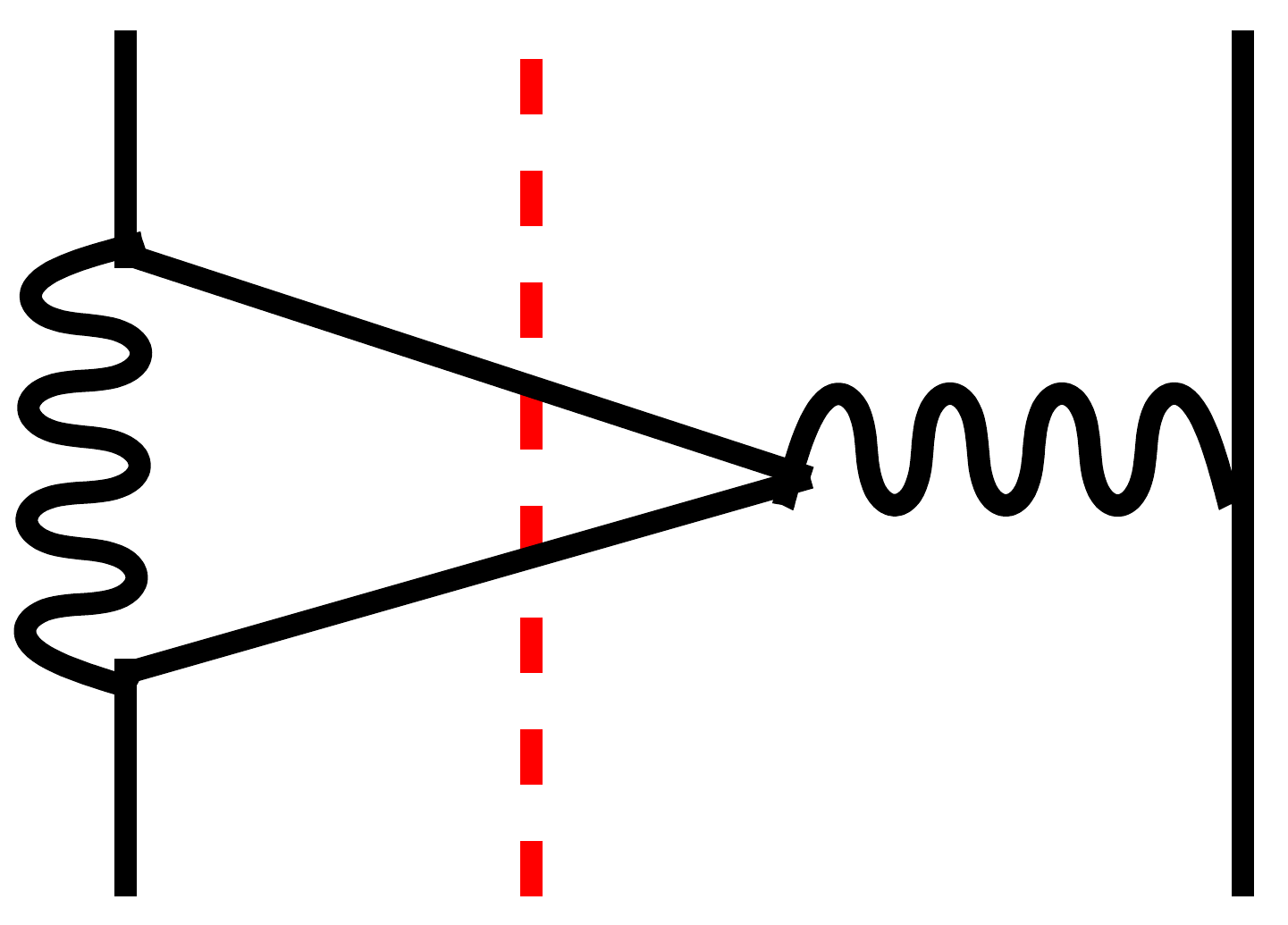}
  \label{subfig:Yf}
}
\subfigure[]{
  \includegraphics[height = 2.5cm]{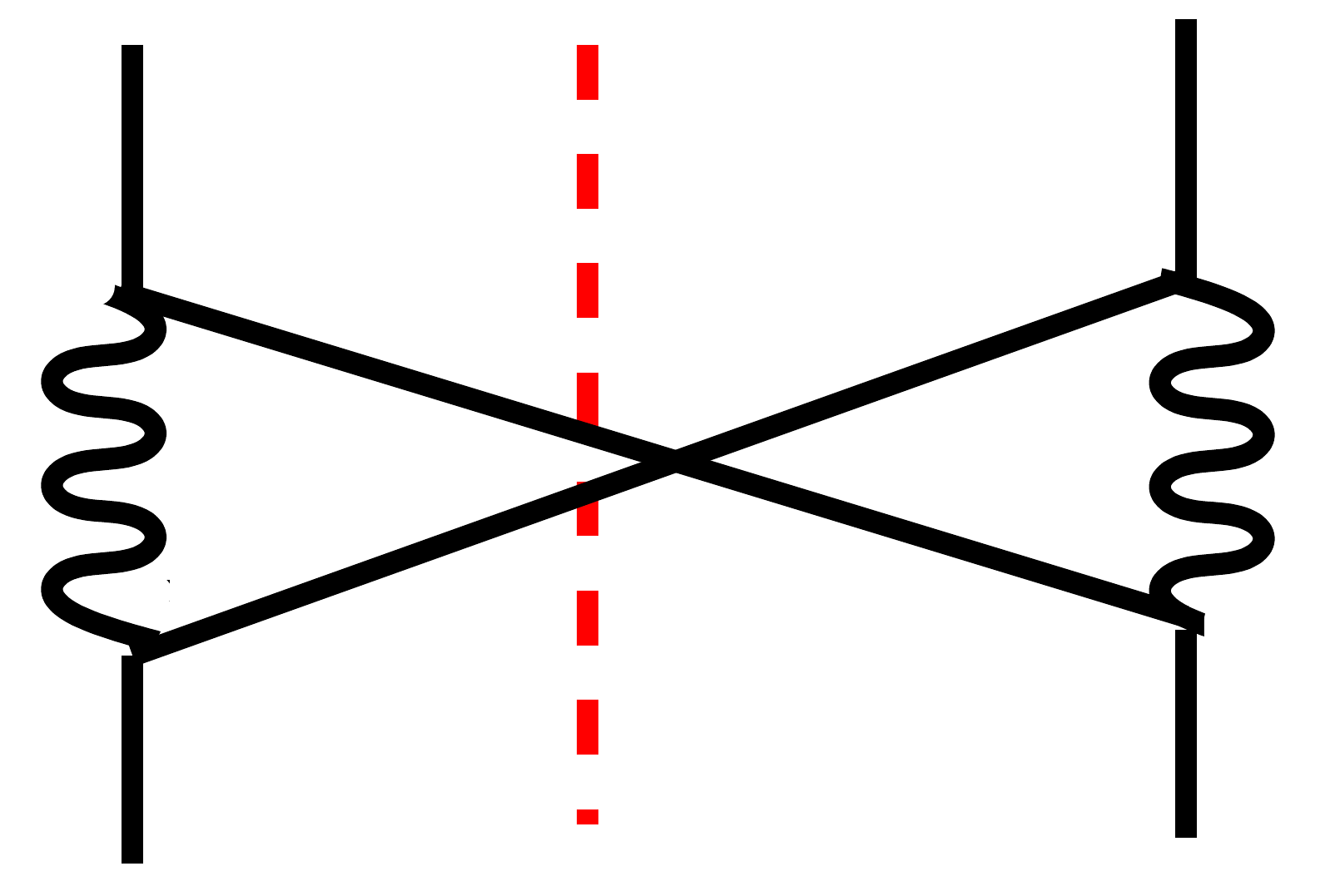}
  \label{subfig:Xf} 
}
\caption{Real contributions to NLO non-singlet DGLAP kernel.}
\label{fig:diagrams}
\end{centering}
\end{figure}

Our main object of interest in the present
work are the diagrams depicted in
Fig.~\ref{fig:diagrams},
with two emitted  on-shell quarks and/or gluons,
that is diagrams with two cut lines.
We will call them 2-real, or shortly 2R, contributions.
The other diagrams with 1-real and 1-virtual, nicknamed 1V1R,
will be also partly considered.
Diagrams with 2 virtual (2V) will not be discussed, because
they will be treated in the same way as in CFP
(deduced from the baryon number conservation rule, which we
keep in the Monte Carlo by construction).

The NLO 2PI diagrams feature amplitude-squares, depicted
in the upper row in Fig.~\ref{fig:diagrams}:
double gluon emission~\ref{subfig:Br},
gluon pair production~\ref{subfig:Vg}
and fermion pair production~\ref{subfig:Vf} as well as interferences,
displayed in the lower row in Fig.~\ref{fig:diagrams}.
Interference diagrams enter into the MC kernels for the first time
at NLO and their correct incorporation in the MC requires 
more care. They can potentially make negative contributions
to the kernel, spoiling the MC weight. It turns out that
the interference diagrams have different statuses in the MC, in
the following we explain the reasons for it.

Let us make a simple observation that 
interferences \ref{subfig:Yg} and \ref{subfig:Bx} together with squares
\ref{subfig:Br} and \ref{subfig:Vg} form the full amplitude square
\[
 \Bigg|~~ 
   \raisebox{-11pt}{\includegraphics[height=11mm]{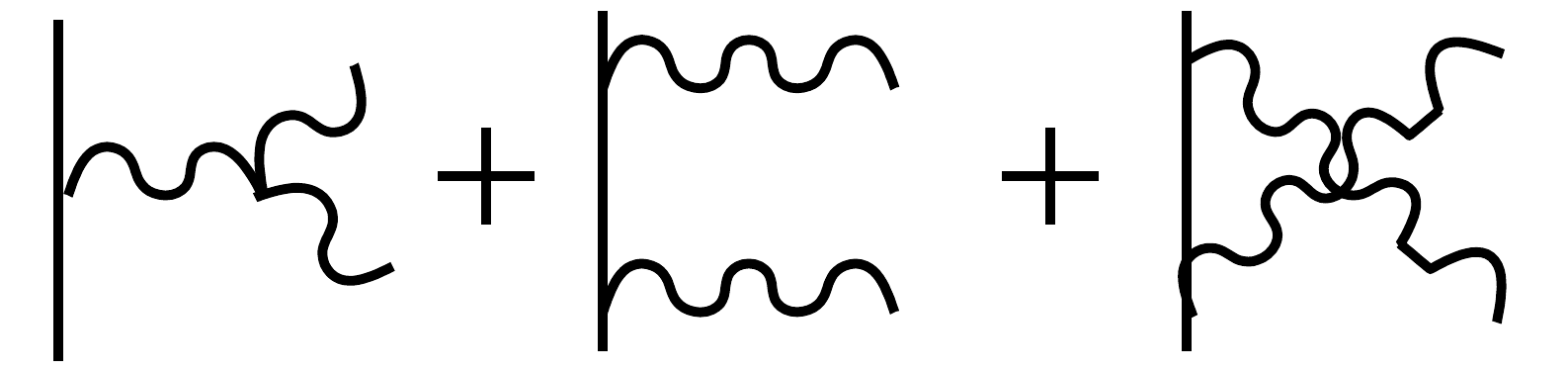}}
 \Bigg|^2.
\]
Positiveness of this amplitude square
will cause the MC weight introducing these two interferences
to be positive. Both interferences~\ref{subfig:Bx} and~\ref{subfig:Yg}
can be implemented in the MC in the 2R group of diagrams.

The interference diagram~\ref{subfig:Yf} originates from the following
amplitude-squared
\[
 \Bigg|~~ 
  \raisebox{-11pt}{\includegraphics[height=11mm]{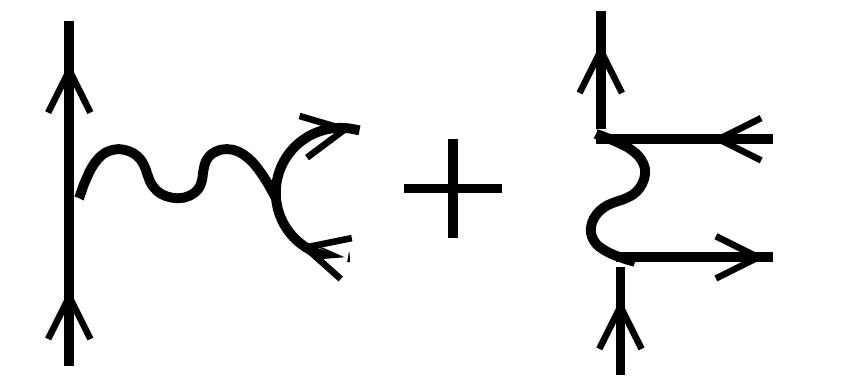}}
 \Bigg|^2,
\]
where $q\bar q$ pair production \ref{subfig:Vf} 
is already included in the non-singlet class,
while $q\bar q$ transitions amplitude (squared)
belongs to the singlet kernel.
Apparently, diagram~\ref{subfig:Yf} corrects quark-gluon transitions absent in  
non-singlet kernel and must be included in the MC 
together with the singlet contributions.

Similarly, the diagram~\ref{subfig:Xf}
is the interference part in the amplitude square
\[
 \Bigg|~~ 
 \raisebox{-11pt}{\includegraphics[height=11mm]{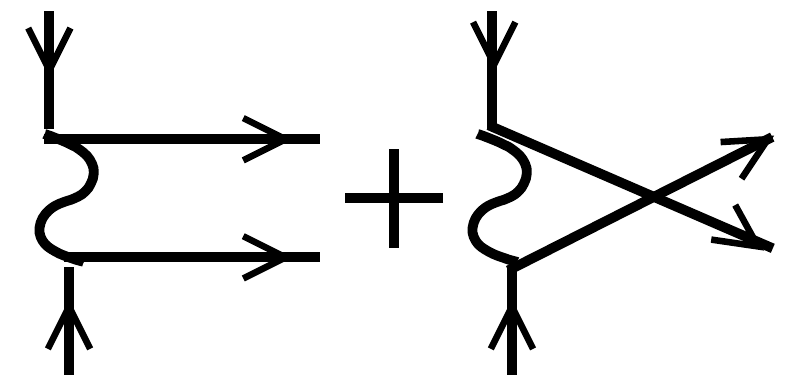}}
 \Bigg|^2,
\]
where both amplitudes representing  $q\bar q$
transitions essentially belong to the singlet NLO kernel.

In the case of both interferences related to quark-gluon transitions
the MC weight is not protected by the Schwartz inequality due to
the absence of amplitude squares in the non-singlet kernels.
The above problem will naturally disappear when singlet diagrams
are included in the game, and we need some temporary fix,
while staying in the non-singlet class.
In order to keep the fermion number conservation we keep these
contributions in the kernel, but add them to the 1R1V class
and treat inclusively.
In the following such a combination of the 1R1V
and integrated subclass of 2R interferences we will call collectively
{\em unresolved} contribution.

Another important point concerns internal singularities
of Feynman diagrams. They are present in graphs of
Fig.~\ref{subfig:Br}, \ref{subfig:Vg} and~\ref{subfig:Vf}.
The double bremsstrahlung diagram Fig.~\ref{subfig:Br}
does not enter into NLO kernel as a whole%
\footnote{ It is not two-particle-irreducible (2PI)}
but only what remains after subtracting
{\em soft collinear counterterm} of the CFP factorization.
On the other hand, diagram of Fig.~\ref{subfig:Vg}
features internal collinear singularity cancelled by the
corresponding virtual (gluon self-energy) diagram.
These diagrams may
enter into the {\em unresolved} part in the MC,
or may be modelled in an exclusive manner.
In the latter case diagram~\ref{subfig:Vg} requires the
construction of a dedicated soft-collinear counterterm
which encapsulates the above internal singularities
and is instrumental in the MC construction.
Such a counterterm will be defined
in Section~\ref{sec:Vg_res}.
Diagram~\ref{subfig:Vf} can be also modelled in an exclusive way
and also needs a dedicated counterterm, see Section~\ref{sec:Vf_res}.

Before that, Section~\ref{sec:LO} presents an overview
of calculations illustrated by the example of the LO kernel.
Notation and methodology
of extracting kernels will be introduced using this simple example.
The integration procedure for NLO kernels is reviewed in
Sections~\ref{sec:res} and~\ref{sec:unres}.
The contributions from each appropriate Feynman diagram
in the axial gauge are calculated in integrated
and unintegrated form needed for the MC.
Analytical integration for control will also be done.
Discussion of the collinear and soft singularity structure
will have the highest priority.

Sections~\ref{sec:res} and \ref{sec:unres}
summarize results of analytical integrations.
In contrast to the results presented in~\cite{Curci:1980uw},
2R contributions will be shown separately instead of the sum of 2R and 1R1V.
Section~\ref{sec:conclusios} provides final discussions and conclusions.

%\vfill\newpage
%%%%%%%%%%%%%%%%%%%%%%%%%%%%%5
\section{Leading order example}
%%%%%%%%%%%%%%%%%%%%%%%%%%%%%5
\label{sec:LO}
\begin{figure}[hhh]
  \begin{center}
    \includegraphics[width=4cm]{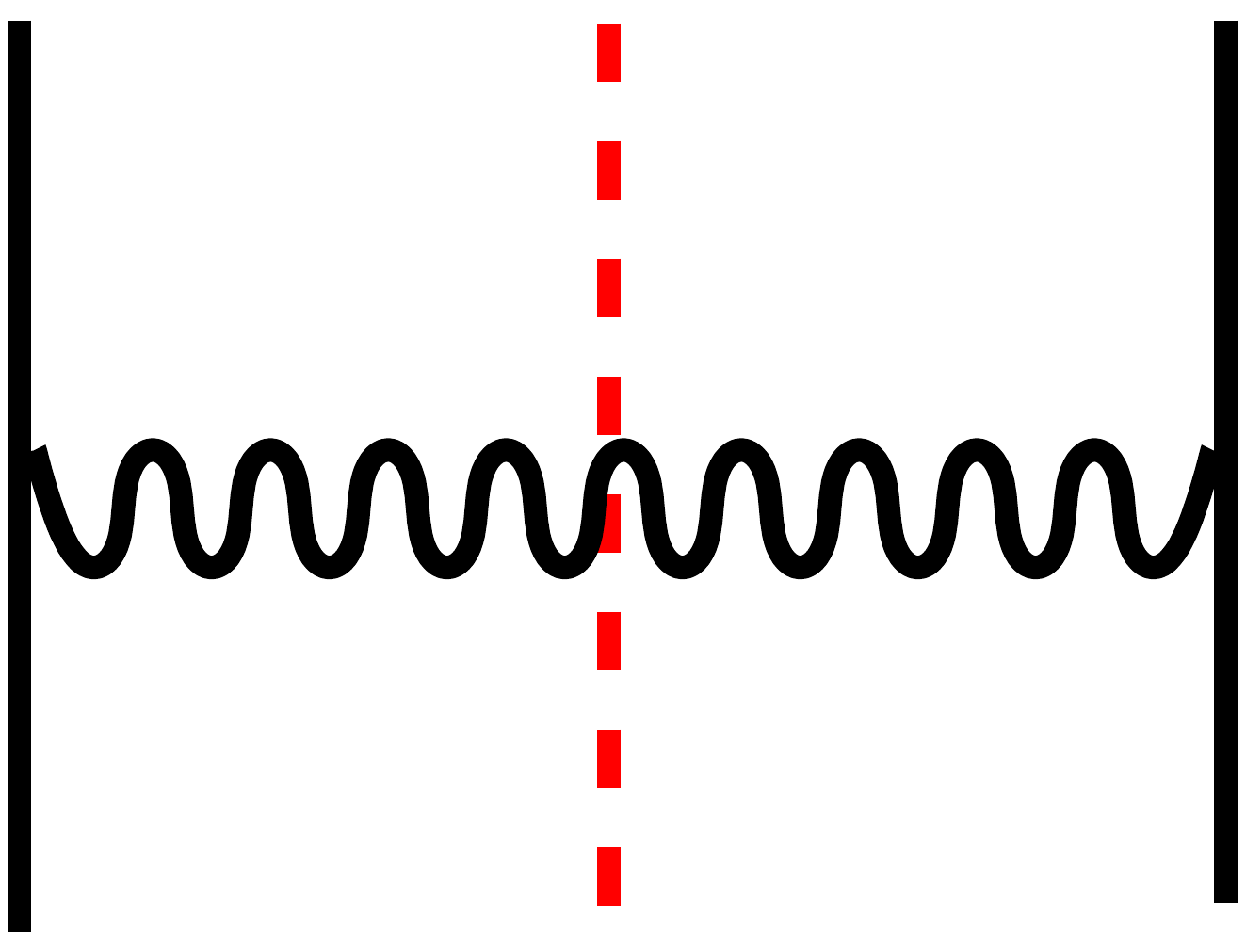}
  \end{center}
\caption{ Born level diagram.}
\label{fig:born}
\end{figure}
Using the LO diagram of Fig.~\ref{fig:born}
we are going to introduce some basic notation,
which will also be useful in the NLO calculations.
In addition we will show correspondence between elements of
the CFP~\cite{Curci:1980uw} scheme
using dimensional regularization ($ n=4+2\epsilon$)
and calculations in the Monte Carlo ($n=4$)
in this simple example.

On one hand, the non-singlet bare PDF of CFP~\cite{Curci:1980uw},
from which the DGLAP kernel is extracted, reads as follows
\begin{equation}
\Gamma^{LO}(x,\epsilon) = 
   \delta(1-x) (1- Z_F^{[1]})
+\text{PP}
 \bigg[
   \int d\Psi_n(k)\; \mu^{-2\epsilon}\;
   x\delta\Big(x-\frac{qn}{pn}\Big)\;
   C_Fg^4 \;
  W^{LO}(k,\epsilon)
 \bigg],
\label{eq:diag}
\end{equation}
where PP is the pole part operator, 
$C_F$ is the color factor,
$Z_F=1+Z^{[1]}+\dots$ is the fermion wave function renormalization factor.

Let us explain step by step the other elements in the above formula.
One particle phase space of emitted on-shell gluon (cut line),
in $n=4+2\epsilon$ dimensions, reads:
\begin{equation}
d\Psi_n(k) = \frac{d^nk}{(2\pi)^n}(2\pi)\delta^+(k^2)
      = \frac{1}{2(2\pi)^{3+2\epsilon}}
        d\alpha\alpha^{1+2\epsilon}
        d|\ba||\ba|^{1+2\epsilon}
        d\Omega_{2+2\epsilon}.
\end{equation}
The emitted gluon 4-momentum $k^\mu$ is parametrized
using Sudakov variables
\begin{equation}
  k = \alpha p + \beta n + k_{\perp},
\label{eq:sdakov_par}
\end{equation}
where $p$ is the momentum of the incoming parton 
and the lightlike vector $n$ defines axial gauge.
The conditions $p^2=k^2=0$ lead
to relation $\beta=-\frac{k_{\perp}^2}{2\alpha(pn)}$.
Non-abelian coherence effects in the soft limit
beyond LO are easier to handle if we use the variable
\begin{equation}
  \ba = \frac{\bk}{\alpha},
\end{equation}
instead of transverse momentum $\bk$.
Its modulus, $|\ba|\equiv a$, will be used to define the
factorization scale in the MC -- we will refer to it as
an {\em angular scale}.%
\footnote{
 Traditional rapidity variable is equal, up to a constant,
 to its logarithm, $\eta=\frac{\ln(k^+/k^-)}{2} =\ln|\ba|+const$.}

Phase space in eq.~\eqref{eq:diag} requires
to be closed from the above, at least temporarily.
In the CFP scheme this closure plays a marginal role,
as $\Gamma(x,\epsilon)$ consists of pure poles.
The upper limit merely influences intermediate results,
through the parametrization of the phase space,
before taking PP and executing all kind
of internal infrared (IR) cancellations.
On the contrary in the Monte Carlo scheme the variable defining
the upper limit of the phase space plays an important role of the
factorization scale.
Its logarithm is the evolution time variable in the MC algorithm/code.
The most popular choices are:
maximum angular scale (angular ordering),
and maximum transverse momentum of all real emitted partons.
Virtuality of the emitter line
in the ladder or maximum $k^-=k^0-k^3$ are the other valid
but less attractive choices.

The function $W$ is just one simple $\gamma$-trace factor 
(see~\cite{Curci:1980uw}):
\begin{equation}
\begin{split}
W^{LO}(k,\epsilon) &= \frac{1}{4(qn)}\frac{1}{q^4}
    \text{Tr}[\slashed{p}\gamma^{\alpha}\slashed{q}
    \slashed{n}\slashed{q}\gamma^{\beta}] d_{\alpha\beta}(k)
%\\&
   = \frac{2}{x}\;
     \frac{1+(1-\alpha)^2+\epsilon\alpha^2}{\alpha^2}
     \frac{1}{\ba^2}.
\end{split}
\end{equation}
Setting the upper phase space limit for the angular variable,
$|\ba|\le Q$, we obtain:%
\footnote{We will often use shorthand 
  notation $\delta_{a=b}\equiv\delta(a-b)$,
  $\Omega_{2+2\epsilon}=\frac{2\pi^{1+\epsilon}}{\Gamma(1+\epsilon)}$
  and $\alpha_S=g^2/(4\pi)$.
}
\begin{equation}
\begin{split}
&\Gamma^{LO}(x,\epsilon)= 
  \delta_{x=1}(1+Z_F^{[1]})
 +\text{PP}\bigg\{
  \frac{C_F}{\mu^{2\epsilon}}
  \frac{g^2}{(2\pi)^{3+2\epsilon}} \int d\Omega_{2+2\epsilon}
  \int d\alpha\; \delta_{1-x=\alpha}
\\&~~~~~~~~~\times
   \frac{2-2\alpha+(1+\epsilon)\alpha^2}{\alpha^{1-2\epsilon}}
   \int_0^{\infty} d|\ba||\ba|^{2\epsilon-1} \;\theta_{|\ba|\le Q}
  \bigg\}
\\&
=\delta_{x=1}(1+Z_F^{[1]})
+\text{PP}
 \bigg\{
  \frac{g^2C_F}{\mu^{2\epsilon}}
   \frac{\Omega_{2+2\epsilon}}{(2\pi)^{3+2\epsilon}}
   \frac{1+x^2 -\epsilon (1-x)^2}{(1-x)^{1-2\epsilon}}
   \frac{Q^{2\epsilon}}{2\epsilon}
\bigg\}
\\
&=\delta_{x=1}
  +\frac{1}{2\epsilon}\;
   \frac{2C_F\alpha_S}{\pi} 
   \bigg(\frac{1+x^2}{2(1-x)} \bigg)_+.
\end{split}
\end{equation}
In the above 
$Z_F^{[1]}=-\frac{1}{\epsilon} \frac{2C_F\alpha_S}{\pi}
  \big(\ln\frac{1}{\delta} -\frac{3}{4} \big)$
provides proper normalization,
with regularization of the IR pole
$\frac{1}{1-x}\to \frac{1-x}{(1-x)^2+\delta^2}$
done exactly as in CFP.
The evolution kernel is defined as twice the residue
of $\Gamma$ at $\epsilon=0$:
\begin{equation}
\label{eq:kerLOres}
P^{LO}_{qq}(\alpha_S,x)=2\text{Res}_0\big(\Gamma^{LO}(x,\epsilon)\big)
=  \frac{2C_F\alpha_S}{\pi} 
   \bigg(\frac{1+x^2}{2(1-x)} \bigg)_+.
\end{equation}

How does the above compare with the Monte Carlo?
In the Monte Carlo the same integral
taken in the limits $q_0<a<Q$ in $n=4$ looks simpler:
\begin{equation}
\begin{split}
&G^{LO}(x,Q)
=(1-S_1^{[1]})\delta_{x=1}+
\int d\Psi_4 xC_Fg^4 W^{LO}(k,\epsilon=0) 
\delta_{x=1-\alpha} \theta_{Q>a>a_0}
\\&
=(1-S_1^{[1]})\delta_{x=1}
%\\&~~~
+\frac{2C_F\alpha_S}{\pi^2}\!\!\!\!
 \int\limits_{Q>a>q_0} \!\!\! \frac{d^3k}{2k_0}
 \frac{1+x^2}{2 \bk^2}
 \delta_{x=1-\alpha}
 \theta_{\alpha>\delta}
%\\&~~~~~~~~~~
 = \ln\frac{Q}{q_0}\;
   \frac{2C_F\alpha_S}{\pi}
   \bigg( \frac{1+x^2}{2(1-x)} \bigg)_+,
\end{split}
\end{equation}
where
$S_1^{[1]}=\ln\frac{Q}{q_0} \frac{2C_F\alpha_S}{\pi}
  \big(\ln\frac{1}{\delta} -\frac{3}{4} \big)$
is the Sudakov formfactor.
As we see, the same LO kernel is now
the coefficient in front of the collinear logarithm:
\begin{equation}
\label{eq:kerLOlog}
\Peu^{LO}_{qq}(\alpha_S,x)=
\frac{\partial}{\partial \ln Q}
 G^{LO}(x,Q)
=  \frac{2C_F\alpha_S}{\pi} 
   \bigg(\frac{1+x^2}{2(1-x)} \bigg)_+.
\end{equation}
Apparently, $1/\epsilon$ pole of CFP translates
into $\ln\frac{Q}{q_0}$:
\begin{equation}
\int_0^Q d|\ba||\ba|^{2\epsilon-1}
   = \frac{Q^{2\epsilon}}{2\epsilon}
     \rightarrow \int_{q_0}^{Q} \frac{d|\ba|}{|\ba|}
   = \ln\frac{Q}{q_0}.
\end{equation}
The relation between PDFs and evolution kernels
in CFP and Monte Carlo factorization schemes 
is, however, more complicated beyond LO.
This is discussed in more detail in ref.~\cite{Kusina:2011xh}, see also
refs.~\cite{Kusina:2010gp,Jadach:2011cr}.
Generally, all differences between MC and CFP factorization
schemes will be traced back to diagrams with subtractions,
or with internal collinear singularity cancellations.

%\vfill\newpage
%%%%%%%%%%%%%%%%%%%%%%%%%%%%%5
%%%%%%%%%%%%%%%%%%%%%%%%%%%%%5
\section{2R contributions to non-singlet NLO kernels}
%%%%%%%%%%%%%%%%%%%%%%%%%%%%%5
\label{sec:res}

In the following we shall calculate
bare PDFs and the resulting inclusive
evolution kernels from 2-real phase space of the
Feynman diagrams contributing to the
non-singlet NLO DGLAP kernels in QCD.
Let us stress that
our real aim are the distributions over the 2R phase
space integration.
Analytical integrations will be performed
mainly as a crosscheck with the know results and for testing
parts of the Monte Carlo code.
% and sometimes in order to deduce 1R1V contributions
% (needed for the MC) from the known total 2R+1R1V,
% instead or calculating them.

We shall start with explaining notation 
and 2R phase space parametrization used
in the calculations.
Differential distributions
and 2R phase space integrals will
be listed for each Feynman diagram separately.

%%%%%%%%%%%%%%%%%%%%%%%%%%%%%5
\subsection{Kinematics}
%%%%%%%%%%%%%%%%%%%%%%%%%%%%%5

Sudakov parametrization is introduced for both emitted partons:
\begin{equation}
k_i = \alpha_i p + \beta_i n + k_{i\perp}
\quad
q_i = p-k_i% = x_i p + x_i^- n + q_{i\perp}
\quad \text{for }i= 1, 2
\label{eq:parametrization}
\end{equation}
where $p^\mu$ is the momentum of the incoming quark
($p^2=0$) and lightlike $n^\mu$ is the axial gauge vector.
Real on-shell emitted gluon or quark has 4-momentum $k_i^\mu$,
and $q_i^\mu$ denotes 4-momentum of the virtual
(off-shell) emitter parton.
We will also denote
$
k = k_1 + k_2,\quad
q = p - k.
$
4-dimensional transverse momenta
$k_{i\perp} = (0, \bk_i, 0)$ for $i=1,2$
will be also used
(to be extended to
$n=4+2\epsilon$ dimensions wherever necessary).
From $k_1^2 = k_2^2 = 0$ we obtain
\begin{equation}
\beta_i = \frac{\bk_i^2}{2\alpha_i(p\cdot n)}.
\end{equation}
The lightcone variable decreases from 1 to
$x = 1 - \alpha_1 - \alpha_2$ after two emissions.
The angular scale variable
\begin{equation}
\ba_i = \frac{\bk_i}{\alpha_i}
\end{equation}
is a preferred choice, instead of transverse momentum.
The virtuality of the emitter parton after two emissions
(entering its propagator) and the gluon pair effective mass are:
\begin{equation}
\begin{split}
&
q^2=
-\alpha_1\alpha_2 \tilde{q}^2(\ba_1,\ba_2),\quad
\tilde{q}^2(\ba_1,\ba_2)
=    \frac{1-\alpha_2}{\alpha_2}\ba_1^2
   + \frac{1-\alpha_1}{\alpha_1}\ba_2^2 + 2\ba_1\cdot\ba_2,
\\&
k^2=\alpha_1\alpha_2 \ba^2(\ba_1,\ba_2),\quad
\ba^2(\ba_1,\ba_2) = \ba_1^2 + \ba_2^2 - 2\ba_1\cdot\ba_2.
\end{split}
\end{equation}

\subsection{Inclusive evolution kernels, CFP vs. MC}
%%%%%%%%%%%%%%%%%%%%%%%%%%%%%%%%%%%%%%%%%%%%%%%%%%%%%

In the CFP scheme, the NLO inclusive kernel is extracted from the second
order expression for the bare PDF,
which in compact CFP notation reads:
\begin{equation}
\label{eq:EGMPR}
\begin{split}
&\Gamma 
= Z_F \frac{1}{\Ibbm-\Pbbm K_0 (1-(1-\Pbbm)K_0)^{-1}}
= \Ibbm Z_F^{(2)}
 +(1+Z_F^{[1]})\Pbbm K_0^{[1]}
\\&~~~~~~~~~~~
 +\Pbbm K_0^{[2]}
 +\Pbbm K_0^{[1]} ((\Ibbm-\Pbbm)K_0^{[1]})
 +(\Pbbm K_0^{[1]})(\Pbbm K_0^{[1]})
 +{\cal O}(\alpha_S^3)
\end{split}
\end{equation}
where $Z_F^{(2)}=1+Z_F^{[1]}+Z_F^{[2]}$
is the quark renormalization constant
and $K_0= K_0^{[1]}+K_0^{[2]}$
is the 2-particle irreducible kernel (truncated to 2-nd order
in perturbative expansion)
defined in refs.~\cite{Ellis:1978sf,Curci:1980uw}.
The NLO contributions to the evolution kernels are coming from
$\Pbbm K_0^{[2]}$
and $\Pbbm K_0^{[1]} ((\Ibbm-\Pbbm)K_0^{[1]})$.
Diagrams (b-g) in Fig.~1
are in the first class and only
diagram (a) is in the second class (with subtraction).
The other 2-nd order terms like $(\Pbbm K_0^{[1]})(\Pbbm K_0^{[1]})$
and  $Z_F^{[1]} \Pbbm K_0^{[1]}$ yield pure $\frac{1}{\epsilon^2}$
poles times elements of LO kernels
and do not contribute to NLO kernels.
The 1-st order $\Ibbm Z_F^{[1]} +\Pbbm K_0^{[1]}$
was already analyzed in the previous section.
From now on we drop flavor indices as all but one diagram contributing
to the non-singlet kernel at NLO level describes $q q$ transitions.
The $q q$  flavor
indices will be understood implicitly if not indicated otherwise.%
\footnote{Possible ways of implementing multi-flavor partons in MC have been presented in
refs. \cite{GolecBiernat:2006xw, Jadach:2007qa}.}
The NLO contribution to the evolution kernel
(to bare PDF of CFP)
from a given 2R Feynman diagram $X$
in Fig.~1 reads:
\begin{equation}
\label{eq:KerNLOcfp}
\begin{split}
&P^X(x)= 2{\rm Res}_0 \big(\Gamma^X(x,\epsilon)\big),
\\&
\Gamma^X(x,\epsilon)
 = {\rm PP} \bigg\{ \frac{1}{\mu^{4\epsilon}} \int d\Psi_n(k_1) d\Psi_n(k_2)\;
     x\delta\Big(x-\frac{qn}{pn}\Big)\;
     C g^4 W^X(k_1,k_2,\epsilon)\;
     \theta_{Q>s(k_1,k_2)} \bigg\},
\end{split}
\end{equation}
where the $\theta$-function limits phase space from the above
using variable $s(k_1,k_2)=\max\{a_1,a_2\}$
(resulting CFP kernel is independent of this cut-off), $C$ is
a color factor of a diagram $X$.

Monte Carlo featuring complete NLO evolution,
can be expressed as a time-ordered exponential
in the logarithm of its factorization scale $Q$,
see~\cite{Jadach:2010ew}.
Hence, at the inclusive level,
it obeys its own evolution equation in $\ln Q$
with its own inclusive NLO
evolution kernel, being the derivative
in $\ln Q$ of the MC~\cite{Jadach:2010ew} distribution
(truncated to 2-nd order)%
\footnote{Strictly speaking, diagrams
  like (b-c) in Fig.~\ref{fig:diagrams} produce
  $\ln^k(Q/\mu_R)$, $k>1$ 
  (similarly to higher $1/\epsilon^k$ poles in CFP),
  to be resummed into running coupling constant,
  before applying this formula.}
\begin{equation}
%\Peu(x)=
  \frac{\partial}{\partial \ln Q}
  \int\! d{\rm Lips}\; 
  \delta_{1-x=\sum\alpha_i}\;
  \Pbbm'_Q \big\{^s K_0 \cdot (1-\Pbbm'_s) K_0 \big\}.
\end{equation}
Here $K_0$ is the same as in CFP and comes from the Feynman diagrams,
albeit with real-virtual collinear cancellations executed before
taking the derivative --
so above formula is finally executed in $n=4$.

For the use of $\Pbbm'$ it is enough to apply
eqs.~(\ref{eq:KerNLOmcb},\ref{eq:KerNLOmca}) below.
It acts on the integrand,
contrary to $\Pbbm$ of CFP acting on the integrals,
hence it provides unintegrated NLO
distributions for the 
MC~\cite{Jadach:2010ew,Jadach:2011cr}. 
For our purpose (2R diagrams) the above reduces to
the following:
\begin{equation}
\label{eq:KerNLOmc}
\Peu(x)=
  \frac{\partial}{\partial \ln Q}
  (G_b(Q,x)+G_a(Q,x)),
\end{equation}
where
\begin{equation}
\label{eq:KerNLOmcb}
G_b(Q,x)=
  x \int\! 
  d\Psi_4(k_1) d\Psi_4(k_2)\;
  \delta_{1-x=\alpha_1+\alpha_2}\;
   \Pbbm' K_0^{[2]}\;
   \theta_{Q>s(k_1,k_2)>q_0}
\end{equation}
contains two-particle-irreducible diagrams
and
\begin{equation}
\label{eq:KerNLOmca}
\begin{split}
&G_a(Q,x)=
  x \int\! 
  d\Psi_4(k_1) d\Psi_4(k_2)\;
  \delta_{1-x=\alpha_1+\alpha_2}\;
\\&~~~~~~~~~~~~
  \times\bigg\{
   \Pbbm' ( K_0^{[1]} K_0^{[1]})
   \theta_{Q>s(k_1,k_2)>q_0}\;
  -\Pbbm'(K_0^{[1]}) \Pbbm'(K_0^{[1]})
   \theta_{Q>s(k_1)>s(k_2)>q_0}
   \bigg\}
\end{split}
\end{equation}
contains diagrams requiring soft counterterms.%
\footnote{Cut-off $q_0$ plays no role in evolution kernel.}
The remaining action of $\Pbbm'$ is spin projection,
the same way as in CFP.
% For the purpose of present work $\Pbbm' K_0^{[2]}$
% and $\Pbbm' ( K_0^{[1]} K_0^{[1]})$ are
% read off from Feynman diagrams, are the same as in CFP.
On the other hand,
the subtraction term $-\Pbbm'(K_0^{[1]}) \Pbbm'(K_0^{[1]})$
is identical to the double gluonstrahlung LO distribution of the MC,
hence it deviates from CFP.
For example interference diagrams contribute
\begin{equation}
\label{eq:Precipe1}
G^X_b(Q,x)=
  \int\! d\Psi_4(k_1) d\Psi_4(k_2)\;
  \delta_{1-x=\alpha_1+\alpha_2}\;
  xC g^4 W^X(k_1,k_2,0)\;
  \theta_{Q>s(k_1,k_2)>q_0}\;
\end{equation}
and subtracted NLO diagrams contribute
\begin{equation}
\label{eq:Precipe2}
\begin{split}
&
G^X_a(Q,x)=
  \int\! d\Psi_4(k_1) d\Psi_4(k_2)\;
  \delta_{1-x=\alpha_1+\alpha_2}\;
  xC g^4
\\&~~~~~~~~~~~~~
\times\bigg[
  W^X(k_1,k_2,0)
  \theta_{Q>s(k_1,k_2)>q_0}\;
 -W^{ct}(k_1)W^{ct}(k_2)
  \theta_{Q>s(k_1)>s(k_2)>q_0}
\bigg].
\end{split}
\end{equation}
Since MC distribution $W^{ct}(k_1)W^{ct}(k_2)$
encapsulates (by construction)
all collinear and soft singularities,
subtracted $G^X_a(Q,x)$ can be evaluated in $n=4$.

Alternative expressions for NLO
inclusive kernels of eq.~(\ref{eq:KerNLOcfp})
and eq.~(\ref{eq:KerNLOmc}) provide precisely
the same results
for graphs (d-e) in Fig.~\ref{fig:diagrams},
which do not have internal
divergences nor require subtraction (as in LO case), 
and are evaluated at $n=4$.
For diagrams (a-c) in Fig.~\ref{fig:diagrams} we shall see
certain small but important differences,
which indicate that the MC represents a slightly different
factorization scheme than CFP.

\subsection{Overview of the 2R phase space integration}
%%%%%%%%%%%%%%%%%%%%%%%%%%%%%%%%%%%%%%%%%%%%%%%%%%%%%%%%%%%%%%%%
\label{sec:integr}

It is convenient to introduce slightly differently
normalized phase space
\begin{equation}
d\Phi_{n}(k)
= \frac{d^{n-1} k}{2k_0}\frac{1}{|\bk|^2}
= \frac{d\alpha}{2\alpha}
  \frac{d |\bk|}{|\bk|} 
  |\bk|^{2\epsilon}
  d\Omega_{2+2\epsilon}
= \frac{d\alpha}{2\alpha}
  \frac{d a}{a} 
  (\alpha a)^{2\epsilon}
  d\Omega_{2+2\epsilon}
= d\Psi_n(k)\frac{(2\pi)^{3+2\epsilon}}{a^2 \alpha^2},
\end{equation}
which is dimensionless in $n=4$.

Most of the presented differential results will
be normalized using the above integration element.
For instance in eq.~(\ref{eq:KerNLOcfp})
we replace $ d\Psi_{n}(k) \to d\Phi_{n}(k)$ and
\begin{equation}
W^X\to \tilde{W}^X(k_1,k_2,\epsilon) 
= Cg^4x
 \frac{ a_1^2a_2^2 \alpha_1^2 \alpha_2^2 }%
      {(2\pi)^{6+4\epsilon}} W^X
= \frac{Cx}{(2\pi)^{4\epsilon}}\left(\frac{\alpha_S}{2\pi^2}\right)^2
   a_1^2a_2^2 \alpha_1^2 \alpha_2^2 \;
 W^X.
\end{equation}

Let us outline the general methodology
used in the 2R phase space integrations.
It will be described in $n=4$,
with some small modifications
it will also apply in $n=4+2\epsilon$.
The integration procedure consists of the following steps:
%%%%%%%%%%%%%%%%%%%%%%%%%%%%%%%%%%%%%
\begin{enumerate}[(a)]
\item 
Using the identity
$\Theta_{Q>\max\{a_1,a_2\}}\equiv \int_0^Q d\tilde{Q}
  \,\delta_{\tilde{Q}=\max\{a_1,a_2\}}$,
the integration variable $\tilde{Q}$ is introduced:
\begin{equation}
\begin{split}
&G^X
= \int \frac{d\alpha_1}{\alpha_1}
  \frac{d\alpha_2}{\alpha_2}
  \delta_{1-x=\alpha_1+\alpha_2}
  \int\limits_0^Q d\tilde{Q} \;
  \delta_{\tilde{Q}=\max\{a_1,a_2\} }
\\&~~~~~~~~~~~~~\times 
  \int \frac{da_1}{a_1}\frac{da_2}{a_2} \; 
  \frac{1}{2\pi}\int d\phi\;
 \tilde{W}^X(a_1/a_2,\phi,\alpha_1,\alpha_2)
 \;\Theta_{\max\{a_1,a_2\}>q_0}.
\end{split}
\end{equation}
\item 
Dimensionless variables
$y_i=a_i/\tilde{Q} $, $y_i \in [0, 1]$ are introduced:
\begin{equation}
\begin{split}
&G^X = \int \frac{d\alpha_1}{\alpha_1}
 \frac{d\alpha_2}{\alpha_2}
 \delta_{1-x=\alpha_1+\alpha_2}
 \int\limits_{q_0}^Q \frac{d\tilde{Q}}{\tilde{Q}}
 \int_0^1 \frac{dy_1}{y_1}\frac{dy_2}{y_2}\;
 \int \frac{d\phi}{2\pi}
\\&~~~~~~~~~~~~~~\times 
 \tilde{W}^X(y_1/y_2,\phi,\alpha_1,\alpha_2)
 \delta_{1=\max\{y_1,y_2\} }.
\end{split}
\end{equation}
\item
Integration over overall
scale $\tilde{Q}$ is performed:
\begin{equation}
\begin{split}
&G^X
 = \ln\frac{Q}{q_0} \int \frac{d\alpha_1}{\alpha_1}
 \frac{d\alpha_2}{\alpha_2}\delta(1-x-\alpha_1-\alpha_2)
 \int_0^1 \frac{dy_1}{y_1}\frac{dy_2}{y_2}\;
 \int \frac{d\phi}{2\pi}
\\&~~~~~~~~~~~\times 
 \tilde{W}^X(y_1/y_2,\phi,\alpha_1,\alpha_2)
 \delta_{1=\max\{y_1,y_2\} }.
\end{split}
\end{equation}
In $n=4+2\epsilon$ this integration yields
$\int_0^Q d\tilde{Q}
\tilde{Q}^{4\epsilon-1}=\frac{Q^{4\epsilon}}{4\epsilon}$.
\item
Nontrivial azimuthal angle dependence enters
in the kernels only through the relative angle
between 2 partons $\phi=\phi_1-\phi_2$.
Integration over $\phi$ and $\phi_2$ is done.
\item 
Integration over $y_1$, $y_2$
(eliminating $\delta_{1=\max\{y_1,y_2\} }$)
is performed.
\item
Integration over lightcone variables
$\alpha_1$ and $\alpha_2$ 
(eliminating $\delta_{1-x=\alpha_1+\alpha_2}$)
is done using the IR regularization of CFP:%
\footnote{
This leads to two 
elementary integrals:
$
I_0 \equiv \int_0^{1}\frac{d\alpha}{\alpha}
           = -\ln\delta,\quad
I_1 \equiv \int_0^{1}\frac{d\alpha\;\ln\alpha}{\alpha}
           = -\frac{1}{2}\ln^2\delta - \frac{\pi^2}{24},
$
defined as in ref.~\cite{Curci:1980uw}.}
\begin{equation}
\int_0^{1-x}\frac{d\alpha}{\alpha}F(\alpha)
\rightarrow \int_0^{1-x}
\frac{d\alpha \;\alpha}{\alpha^2+\delta^2}F(\alpha).
\end{equation}
\end{enumerate}

%=================================================
\subsection{Gluonstrahlung interference diagram - Bx}
%%%%%%%%%%%%%%%%%%%%%%%%%%%%%%%%%%%%%%%%%%%%%%%%%%%%
Let us start with the relatively simple
ladder interference diagram Bx
of Fig.~\ref{subfig:Bx}.
The expression for 2R dimensionless differential distribution reads
\begin{equation}
\begin{split}
\label{eq:BxW}
\tilde{W}^{Bx}(k_1,k_2)
&= 4\Big(C_F^2-\frac{1}{2}C_AC_F\Big)
    \left(\frac{\alpha_S}{2\pi^2}\right)^2
\\&\times    \frac{\ba_1^2\ba_2^2}{\tilde{q}^4(a_1,a_2)}
    \bigg[T_0^{Bx} + T_1^{Bx}\frac{\ba_1\cdot\ba_2}{\ba_1^2}
    + T_2^{Bx}\frac{\ba_1\cdot\ba_2}{\ba_2^2}
    + T_3^{Bx}\frac{(\ba_1\cdot\ba_2)^2}{\ba_1^2\ba_2^2}\bigg],
\end{split}
\end{equation}
where: %the color factor $C=C_F^2-\frac{1}{2}C_AC_F$ and:
\begin{equation}
\begin{split}
& T_0^{Bx} = 2x\frac{1+x^2}{1-x}\left(\frac{1}{\alpha_1}
           +\frac{1}{\alpha_2}\right) - 2x,\quad
  T_1^{Bx} = \frac{1+2x^2}{\alpha_1} - 1 + x - x^2,\\
& T_2^{Bx} = \frac{1+2x^2}{\alpha_2} - 1 + x - x^2,\quad
  T_3^{Bx} = 2(1+x^2).
\end{split}
\end{equation}
The first order expression for the PDF in the MC
(performing scalar products) reads:
\begin{equation}
\begin{split}
&G^{Bx}(Q,x)
= \int d\Phi_4(k_1)d\Phi_4(k_2)\;
   \delta_{x=1-\alpha_1-\alpha_2}
   \tilde{W}^{Bx}(k_1,k_2)
\\&~~~
= \int \frac{d\alpha_1}{\alpha_1}
   \frac{d\alpha_2}{\alpha_2}
   \delta_{1-x=\alpha_1+\alpha_2}
   \int\limits_0^{\infty}\frac{da_1}{a_1} 
   \int\limits_0^{\infty}\frac{da_2}{a_2}
   \int\limits_0^{2\pi} \frac{d\phi}{2\pi} \;
   4\Big(C_F^2-\frac{1}{2}C_AC_F\Big)
   \left(\frac{\alpha_S}{2\pi}\right)^2
\\&~~~~\times
   \frac{a_1^2a_2^2}{\tilde{q}^4(a_1,a_2)}
   \bigg[T_0^{Bx} + T_1^{Bx}\frac{a_2}{a_1}\cos\phi
       + T_2^{Bx}\frac{a_1}{a_2}\cos\phi + T_3^{Bx}\cos^2\phi
   \bigg]\;
   \theta_{\max\{a_1,a_2\}<Q}.
\end{split}
\end{equation}
The above includes factors $1/2!$ due to Bose-Einstein (BE) symmetrization 
as well as 2 multiplying interference diagrams.
Following the LO calculation example of Section~\ref{sec:integr},
the integration over transverse degrees of freedom is done:
\begin{equation}
\begin{split}
G^{Bx}(Q,x) &= 
 \ln\frac{Q}{q_0}\;
    4\Big(C_F^2-\frac{1}{2}C_AC_F\Big)
    \left(\frac{\alpha_S}{2\pi}\right)^2
    \int \frac{d\alpha_1}{\alpha_1}\frac{d\alpha_2}{\alpha_2}
    \delta_{1-x=\alpha_1+\alpha_2}
\\&\times
 \bigg[ T_0^{Bx}\frac{\alpha_1\alpha_2}{2x}
    - T_1^{Bx}\frac{\alpha_1^2\alpha_2}{2x(1-\alpha_1)}
    - T_2^{Bx}\frac{\alpha_1\alpha_2^2}{2x(1-\alpha_2)}
\\&~~~~~~~~~~~~~~~~~~~~~~
    + T_3^{Bx}\bigg(\frac{1}{4}
    \ln\left(\frac{x}{(1-\alpha_1)(1-\alpha_2)}\right)
    + \frac{\alpha_1\alpha_2}{2x}\bigg)
 \bigg].
\end{split}
\end{equation}
Integration over $\alpha$-variables finally provides:
\begin{equation}
\label{eq:BxP}
\begin{split}
&G^{Bx}(Q,x) = 
  \ln\frac{Q}{q_0}\; \Peu^{Bx}(x),
\\&
% \Peu^{Bx}_{qq}(x)=
%  \Big(\frac{2\alpha_S}{\pi} \Big)^2
%  C\;
%  \frac{1}{8}
%  \bigg[ \frac{1+x^2}{1-x}
%         \Big(4I_0+4\ln(1-x)-\ln^2(x)\Big)
%                + 2(1+x)\ln(x)
%   \bigg].
\Peu^{Bx}(x)=
 \Big(\frac{\alpha_S}{2\pi}\Big)^2
 \Big(C_F^2-\frac{1}{2}C_AC_F\Big)\;
 \bigg[ \frac{1+x^2}{1-x}
        \Big(8I_0+8\ln(1-x)-2\ln^2(x)\Big)
               + 4(1+x)\ln(x)
  \bigg].
\end{split}
\end{equation}
As already said, $\Peu^{Bx}(x)$ is the same
in CFP and in MC schemes (up to a normalization factor 2),%
\footnote{There is a difference between normalization in MC and CFP
kernels at NLO level $\Peu(x)=2P(x)$. It is due to the definition of
MC kernel as a derivative over $\ln Q$ not $\ln Q^2$.}
because this diagram has no internal collinear divergence.
Uncanceled IR divergences are still present ($I_0$ term).

Summarizing, the distribution of eq.~(\ref{eq:BxW})
will enter into the NLO correction to the MC
exclusive kernel contribution. The distribution of eq.~(\ref{eq:BxP})
will be used for numerical overall tests of the MC
at the NLO level.

\subsection{Subtracted double bremsstrahlung diagram}
%%%%%%%%%%%%%%%%%%%%%%%%%%%%%%%%%%%%%%%%%%%%%%%%%%%%%%%%
The double bremsstrahlung diagram of Fig.~\ref{subfig:Br}
(denoted as Br) is not 2PI (it consists of two 2PI LO kernels)
and needs a subtraction term (referred to as diagram BrC).

We shall start with a simpler case of
integrating Br$-$BrC contribution to evolution
kernel in the MC scheme in $n=4$.
Next we shall recalculate the same Br$-$BrC contribution
to the bare PDF and NLO kernel in the CFP scheme,
analyzing all its components and
discussing all differences with the MC case in detail.

The differential distributions for two Br diagrams%
\footnote{ Two diagrams because of interchange of vertices
due to the Bose-Einstein symmetrization.}
in $n=4+2\epsilon$ read as follows:
\begin{equation}
\label{eq:Br12W}
\begin{split}
&\tilde{W}^{Br}(k_1,k_2)
=\tilde{W}^{Br1}(k_1,k_2)+\tilde{W}^{Br1}(k_2,k_1),
\\&
\tilde{W}^{Br1}(k_1,k_2)
= \frac{4C_F^2}{(2\pi)^{4\epsilon}}\left(\frac{\alpha_S}{2\pi^2}\right)^2
    \frac{\ba_1^2\ba_2^2}{\tilde{q}^4(a_1,a_2)}
    \bigg[T_0^{Br} + T_1^{Br}\frac{\ba_1\cdot\ba_2}{\ba_1^2}
    + T_2^{Br}(\epsilon)\frac{\ba_2^2}{\ba_1^2}\bigg],
\end{split}
\end{equation}
where: %the color factor is $C=C_F^2$ and
\begin{equation}
\begin{split}
& T_0^{Br}
    = 1+x^2 + (1-\alpha_1)^2,\quad
  T_1^{Br}
    = 2\frac{1-\alpha_1}{\alpha_1}(1+x^2 + (1-\alpha_1)^2),\\
& T_2^{Br}(\epsilon)= T_2^{Br}(0)+\epsilon T'^{Br}_{2},\quad
  T_2^{Br}(0)= \frac{1}{\alpha_1^2}
      [1+(1-\alpha_1)^2] [x^2+(1-\alpha_1)^2],\\
& T'^{Br}_{2}= \frac{1}{\alpha_1^2}
    [\alpha_1^2(x^2+(1-\alpha_1)^2) +\alpha_2^2(1+(1-\alpha_1)^2)].
\end{split}
\end{equation}
The most singular term $\sim T^{Br}_2$ can be rewritten
(modulo ${\cal O}(\epsilon^2)$ terms)
as a product of two LO kernels:
\begin{equation}
\begin{split}
&T_2^{Br}(\epsilon)=
 \frac{\alpha_2}{\alpha_1} (1-\alpha_1)
 P^{(0)}_{qq}(z_1,\epsilon)
 P^{(0)}_{qq}(z_2,\epsilon),\quad
%      \frac{1+z_1^2 +\epsilon (1-z_1)^2}{1-z_1}
%      \frac{1+z_2^2 +\epsilon (1-z_1)^2}{1-z_2}
\\&
 P^{(0)}_{qq}(z,\epsilon)
   \equiv \frac{1+z^2 +\epsilon (1-z)^2}{1-z}
=P^{(0)}_{qq}(z)+\epsilon P'^{(0)}_{qq}(z),
\end{split}
\end{equation}
where $z_1=1-\alpha_1$ and $z_2=(1-\alpha_1-\alpha_2)/(1-\alpha_1)$.
The above term coincides
in the MC for the ladder with the following counterterm,
being just the LO MC distribution
\begin{equation}
\label{eq:Br12K}
\tilde{K}^{BrC}(k_1,k_2)
=4C_F^2\left(\frac{\alpha_S}{2\pi^2}\right)^2
\frac{\alpha_1^2}{(1-\alpha_1)^2}
 T_2^{Br}(0)
=4C_F^2\left(\frac{\alpha_S}{2\pi^2}\right)^2
\frac{\alpha_1\alpha_2}{1-\alpha_1}
 P^{(0)}_{qq}(z_1)P^{(0)}_{qq}(z_2).
\end{equation}
It also enters as a subtraction term
into the MC weight which implements NLO corrections.
The contribution to the inclusive PDF
of the NLO MC, including the explicit Bose-Einstein
(BE) symmetrization factor $1/2!$, reads:
\begin{equation}
\label{eq:Br12mc}
\begin{split}
&
G^{Br}(Q,x)
=\ln\frac{Q}{q_0}\;
\Peu^{Br}_{sub}(x)
= \frac{1}{2!} \int d\Phi_4(k_1)d\Phi_4(k_2)\;
\delta_{x=1-\alpha_1-\alpha_2}
\tilde{W}^{Br}_{sub} \;\theta_{Q>\max\{a1,a2\}>q_0},
\\&
\tilde{W}^{Br}_{sub}(k_1,k_2)=
 \tilde{W}^{Br1}(k_1,k_2)
     +\tilde{W}^{Br1}(k_2,k_1)
\\&~~~~~~~~~~~~~~~
-\tilde{K}^{BrC}(k_1,k_2)
 \theta_{Q>a_2>a_1}
-\tilde{K}^{BrC}(k_2,k_1)
 \theta_{Q>a_1>a_2},
\end{split}
\end{equation}
where
% \begin{equation}
% \label{eq:PeuSub}
% \begin{split}
% &\Peu^{Br}_{Sub}(x) =
% \left(\frac{2C_F\alpha_S}{\pi}\right)^2
% \frac{1}{8}
% \bigg[
%  \frac{1+x^2}{1-x}
%  \Big(-4I_0-4\ln(1-x)+2\ln^2(x) \Big)
% \\&~~~
%    + 3(1-x)
%    - (1-x)\ln(x)
%    + \frac{1}{2}(1+x)\ln^2(x)
%    - (1+x)\ln(x) 
% \bigg].
% \end{split}
% \end{equation}
\begin{equation}
\label{eq:PeuSub}
\begin{split}
&\Peu^{Br}_{sub}(x) =
\left(\frac{\alpha_S}{2\pi}\right)^2 C_F^2
\bigg[
 \frac{1+x^2}{1-x}
 \Big(-8I_0-8\ln(1-x)+4\ln^2(x) \Big)
\\&~~~
   + 6(1-x)
   - 2(1-x)\ln(x)
   + (1+x)\ln^2(x)
   - 2(1+x)\ln(x)
\bigg],
\end{split}
\end{equation}
is obtained from analytical
phase space integration using the same methodology
as in the CFP scheme, see below.

It should be kept in mind that
$\Peu^{Br}_{sub}(x)$ above
is obtained for the angular ordering
and it would be different if we would have adopted $k^T$-ordering --
the difference would be
$P^{Br}_{Kin}(x)$ of eq.~(\ref{eq:PeuBrKin}) below,
see discussion in refs.~\cite{Kusina:2011xh,Kusina:2010gp}.

Let us now turn to the CFP scheme which provides,
for this particular diagram,
a 2R contribution to the NLO kernel, different than in the MC scheme.
For calculating $\Pbbm K_0^{[1]} ((\Ibbm-\Pbbm)K_0^{[1]})$
of the bare PDF of eq.~(\ref{eq:EGMPR})
we follow procedure (a-f) of
Section~\ref{sec:integr} step by step.
After integrating over $\tilde{Q}$,
$\phi$ and $\phi_2$, at step (e),
we deal in the $\Pbbm (K_0^{[1]} K_0^{[1]})$ part
with the singular integral in the variable
$y=\max(y_1,y_2)$:
\begin{equation}
\label{eq:BrInty}
\bigg(\frac{Q^2}{\mu_F^2}\bigg)^{2\epsilon}
\frac{\Omega^2_{2+2\epsilon}}{4\epsilon}
\int_0^1 \frac{dy}{y^{1-2\epsilon}}
=
\bigg(\frac{Q^2}{\mu_F^2}\bigg)^{2\epsilon}
\frac{\Omega^2_{2+2\epsilon}}{4\epsilon}
\int_0^1 dy \;
\bigg\{ \frac{1}{\epsilon}\delta_{y=0} 
  +\bigg(\frac{1}{y}\bigg)_+
  +{\cal O}(\epsilon^1) \bigg\}.
\end{equation}
The most singular part due to the
$\frac{1}{\epsilon}\delta_{y=0}$ term
in $\Pbbm (K_0^{[1]} K_0^{[1]})$
is easily integrated:
\[
\Gamma^{Br} \simeq
 \bigg[ +\frac{1}{2}\bigg] \frac{1}{\epsilon^2}
  \bigg(\frac{C_F\alpha_S}{\pi}\bigg)^2
  (P^{(0)}_{qq}\otimes P^{(0)}_{qq})(x)\big|_{2R},
\]
where
$(P^{(0)}_{qq}\otimes P^{(0)}_{qq})(x)\big|_{2R}=
 \frac{1+x^2}{1-x} [ 4\ln\frac{1}{\delta} +4\ln(1-x)]
 +(1+x)\ln x -2(1-x)$
is just the double convolution of the LO kernel.
The counterterm $\Pbbm K_0^{[1]} ((-\Pbbm)K_0^{[1]})$
contributes the same expression,
but with $(-1)$ in front,
and finally $(\Pbbm K_0^{[1]})(\Pbbm K_0^{[1]})$
adds the same expression with $(+1)$ in front.
Altogether, the pattern of building correct exponential
structure of the LO in CFP is much more complicated than
in the MC, with a lot of over-subtractions,
see more discussion in  ref.~\cite{Jadach:2010aa}.

Our main aim however is the residue
in front of the $\frac{1}{\epsilon}$ pole.
Most of it comes from the term 
$\big(\frac{1}{y}\big)_+$ in eq.~(\ref{eq:BrInty}),
which is in close correspondence with the NLO correction
in the MC.
In addition it gets ``fall-out'' contributions from
$\frac{1}{\epsilon^2}\times \epsilon$ terms.
In particular $x$-independent terms from the expansion
\[
\frac{1}{(2\pi)^2}\bigg(\frac{Q^2}{\mu_F^2}\bigg)^{2\epsilon}
\frac{\Omega^2_{2+2\epsilon}}{\epsilon}
=\frac{1}{\epsilon}
+2\ln\bigg(\frac{Q^2}{\mu_F^2}\bigg)
+2\omega_2
\]
luckily cancel between $\Pbbm (K_0^{[1]} K_0^{[1]})$ and
the collinear counterterm $\Pbbm K_0^{[1]} ((-\Pbbm)K_0^{[1]})$.
A~similar but $x$-dependent contribution from
$T'^{Br}_2$ gives rise to a remnant term
\[
\frac{1}{4\epsilon}
\int\limits_0^1 dy\; \frac{\delta_{y=0}}{\epsilon} \!\!\!
\int\limits_{x=z_1z_2} \!\!\! dz_1 dz_2
\Big\{
  \frac{1}{2!}[\epsilon P'^{(0)}_{qq}(z_2) P^{(0)}_{qq}(z_1)
 + P^{(0)}_{qq}(z_2) \epsilon P'^{(0)}_{qq}(z_1)]
-[\epsilon P'^{(0)}_{qq}(z_2) P^{(0)}_{qq}(z_1)]
\Big\}.
\]
The second bracket $[\dots]$ comes from the counterterm --
it lacks $\frac{1}{2!}$ of the true distribution,
and one of $P'^{(0)} P^{(0)}$ terms gets killed by the PP operation.
Altogether, the above spin artifact of CFP
(absent in MC) contributes the following:
\begin{equation}
\begin{split}
&
\Gamma^{Br}_{Sp}=
\frac{1}{2\epsilon}
P^{Br}_{Sp}(x),
\\&
P^{Br}_{Sp}(x)=
\left(\frac{\alpha_S}{2\pi}\right)^2C_F^2
\int dz_1 dz_2 \delta_{x=z_1z_2}
\Big[
 - P'^{(0)}_{qq}(z_2) P^{(0)}_{qq}(z_1)
 + P^{(0)}_{qq}(z_2)  P'^{(0)}_{qq}(z_1)
\Big]
\\&~~~~~~~~~
% =\bigg(\frac{2C_F\alpha_S}{\pi}\bigg)^2
%  \frac{1}{8}
%  2(1-x)\ln(x)
=\left(\frac{\alpha_S}{2\pi}\right)^2 C_F^2\;
 2(1-x)\ln(x).
\end{split}
\end{equation}
The last important contribution in the class
$\sim \frac{1}{4\epsilon}\frac{\delta_{y=0}}{\epsilon} \times\epsilon$
is due to the
$(\alpha_1\alpha_2)^{2\epsilon}$ 
term in our particular
choice of the phase space parametrization.
In fact its role is to cancel the dependence on this choice,
see also discussion in ref.~\cite{Kusina:2010gp}.
It is produced by a similar mechanism of partial cancellation
with the counterterm as above, and
in the $z$-parametrization reads:
\[
\begin{split}
&\frac{1}{4\epsilon}
\int\limits_0^1 dy\; \frac{\delta_{y=0}}{\epsilon} \!\!\!
\int\limits_{x=z_1z_2} \!\!\! dz_1 dz_2\;
\Big\{
\frac{1}{2!}[
 2\epsilon\ln(\alpha_1\alpha_2) P^{(0)}_{qq}(z_2) P^{(0)}_{qq}(z_1)]
-[(2\epsilon\ln\alpha_2) P^{(0)}_{qq}(z_2) P^{(0)}_{qq}(z_1)]
\Big\}
\\&~~~~
=\frac{1}{4\epsilon}
\int dz_1 dz_2\;
\delta_{x=z_1z_2}\;
P^{(0)}_{qq}(z_2) P^{(0)}_{qq}(z_1)
\ln\frac{1-z_1}{z_1(1-z_2)}.
\end{split}
\]
Its contribution to the bare PDF and NLO kernel is:
\begin{equation}
\label{eq:PeuBrKin}
\Gamma^{Br}_{Kin}(x)=
\frac{1}{2\epsilon} P^{Br}_{Kin}(x),\quad
% \Peu^{Br}_{Kin}(x)
% =\bigg(\frac{2C_F\alpha_S}{\pi}\bigg)^2
%  \frac{1}{8} 
%  \left( -(1+x)\ln^2(x)+2(1-x)\ln(x) \right).
P^{Br}_{Kin}(x)
=\left(\frac{\alpha_S}{2\pi}\right)^2 C_F^2
 \left[ 2(1-x)\ln(x) - (1+x)\ln^2(x) \right].
\end{equation}

Finally the real physics is in the term 
$\big(\frac{1}{y}\big)_+$ in eq.~(\ref{eq:BrInty}),
which happens to be exactly the same
(up to the normalization factor 2) as the MC contribution
of eq.~(\ref{eq:Br12mc}).
At step (f) of the integration procedure it reads:
\begin{equation}
\begin{split}
&\Gamma^{Br}_{sub} 
= \frac{C_F^2}{\epsilon}
 \left(\frac{\alpha_S}{2\pi}\right)^2
 \int \frac{d\alpha_1}{\alpha_1}
 \frac{d\alpha_2}{\alpha_2}
 \delta_{1-x=\alpha_1+\alpha_2}
\bigg\{ 
  T_0^{Br}\frac{\alpha_1\alpha_2}{2x}
- T_1^{Br}\frac{\alpha_1^2\alpha_2}{2x(1-\alpha_1)}
\\&~~~~~~~~~~~~~~~~~~~~
+ T_2^{Br}
    \frac{\alpha_1^2(\alpha_1\alpha_2-x)}{2x(1-\alpha_1)^2}
+ T_2^{Br}
    \frac{\alpha_1^2}{2(1-\alpha_1)^2}
    \ln\frac{(1-\alpha_1)^2\alpha_2}{x\alpha_1}
\bigg\}.
\end{split}
\end{equation}
After $\alpha$-integrations we obtain
\begin{equation}
\Gamma^{Br}_{sub}
=\frac{1}{2\epsilon} P^{Br}_{sub}(x),
\end{equation}
where $P^{Br}_{Sub}(x)=\frac{1}{2}\Peu^{Br}_{Sub}(x)$ of eq.~(\ref{eq:PeuSub}).

Altogether, the CFP kernel from the subtracted Br diagram is:
% \begin{equation}
% \label{eq:Br12kern}
% \begin{split}
% &\Peu^{Br}_{qq}(x)
% =\Peu^{Br}_{Sub}(x)
% +\Peu^{Br}_{Kin}(x)
% +\Peu^{Br}_{Sp}(x)=
% \\&~~~~
% =\left(\frac{2C_F^2\alpha_S}{\pi}\right)^2 
%  \frac{1}{8}
%  \bigg(
%  \frac{1+x^2}{1-x}[-4I_0-4\ln(1-x)+2\ln^2(x)]
% \\&~~~~~~~~
%   -\frac{1}{2}(1 +x )\ln^2(x)
%   -(1+x)\ln x +3(1-x)\ln x
%   +3(1-x)
% \bigg),
% \end{split}
% \end{equation}
\begin{equation}
\label{eq:Br12kern}
\begin{split}
P^{Br}(x)
&=P^{Br}_{sub}(x)
+P^{Br}_{Kin}(x)
+P^{Br}_{Sp}(x)
\\&
=\left(\frac{\alpha_S}{2\pi}\right)^2 C_F^2
 \bigg(
 \frac{1+x^2}{1-x}[-4I_0-4\ln(1-x)+2\ln^2(x)]
\\&
  -\frac{1}{2}(1+x)\ln^2(x)
  -(1+x)\ln x + 3(1-x)\ln x
  +3(1-x)
\bigg),
\end{split}
\end{equation}
reproducing the result of ref.~\cite{Curci:1980uw}.

As noted in ref.~\cite{Kusina:2010gp}
$P^{Br}_{Kin}(x)$ is absent in CFP, provided we choose
the maximum transverse momentum as factorization
scale variable: $s(k_1,k_2)=\max(k^T_1,k^T_2)$.
It is simply the case because the factor
$ (\alpha_1\alpha_2)^{2\epsilon} $
is absent.
However, an additional contribution
exactly equal to $P^{Br}_{Kin}(x)$ would appear
from the 2R integral of eq.(\ref{eq:Br12mc}) due
to adopting $k^T$-ordering,
and it would exactly compensate the lack of $P^{Br}_{Kin}(x)$
from $(\alpha_1\alpha_2)^{2\epsilon} $.
In a sense,
the CFP scheme features an automatic {\em self-correcting} mechanism,
such that it provides the result for $k^T$-ordering,
independently of the kinematic parametrization of the phase
space actually used.

Summarizing on the subtracted Br diagram,
the integrand $\tilde{W}^{Br}_{sub}$
in eq.~(\ref{eq:Br12mc})
will enter into the NLO correction to the MC distribution,
while eqs.~(\ref{eq:PeuSub}) and~(\ref{eq:Br12kern})
will be used in the numerical tests of the MC codes.
The difference between CFP and MC factorization schemes
for the subtracted Br diagram (for Br+Bx as well)
at the inclusive kernel is IR finite and reads:
% \begin{equation}
% \label{eq:CFPvsMCBr}
% \begin{split}
% &
% \Delta_{\rm CFP-MC} \Peu^{Br}_{sub}(x)=
%  \Peu^{Br}_{Kin}(x)
% +\Peu^{Br}_{Sp}(x)=
% \\&~~~~~~~~~~~~~
% =\bigg(\frac{2C_F\alpha}{\pi}\bigg)^2
%  \frac{1}{8} 
%  \left[ -(1+x)\ln^2(x)+4(1-x)\ln(x)
%  \right].
% \end{split}
% \end{equation}
\begin{equation}
\label{eq:CFPvsMCBr}
\begin{split}
&
\Delta_{\rm CFP-MC} P^{Br}_{sub}(x)=
 P^{Br}_{Kin}(x)
+P^{Br}_{Sp}(x)=
\\&~~~~~~~~~~~~~
=\left(\frac{\alpha_S}{2\pi}\right)^2 C_F^2
 \left[ 4(1-x)\ln(x) - (1+x)\ln^2(x)
 \right].
\end{split}
\end{equation}

Let us also sum up $C_F^2$ contributions of the Br and Bx diagrams
(using only $C_F^2$ part of Bx) in the 2R phase space.
For the MC scheme we obtain:
\begin{equation}
\label{eq:GMCBrX}
\begin{split}
&G^{Br sub + Bx}(Q,x)
= \int d\Phi_4(k_1)d\Phi_4(k_2)\;
\delta_{x=1-\alpha_1-\alpha_2}
( \tilde{W}^{Br}_{sub} +\tilde{W}^{Bx} )\theta_{Q>\max\{a1,a2\}>q_0}\\
&~~~~~~~~~~~~~~~
=\ln\frac{Q}{q_0}\; \Peu^{Br sub + Bx}(x),
\\
% &\Peu^{BrX}_{MC}(x) =
% \Peu^{Br}_{Sub}(x)+\Peu^{Bx}_{qq}(x)
% \\&~~~
% =\left(\frac{2C_F^2\alpha_S}{\pi}\right)^2 
% \frac{1}{8}
% \bigg[
%      \frac{1+x^2}{1-x}\ln^2(x)
%    + (1+x) \Big( \ln(x) + \frac{1}{2}\ln^2(x) \Big)
%    + (1-x) \Big( 3 - \ln(x) \Big)
% \bigg].
&\Peu^{Br sub + Bx}(x) =
\Peu^{Br}_{sub}(x)+\Peu^{Bx}(x)
\\&~~~
=\left(\frac{\alpha_S}{2\pi}\right)^2 C_F^2
\bigg[
     2\frac{1+x^2}{1-x}\ln^2(x)
   + (1+x) \Big( 2\ln(x) + \ln^2(x) \Big)
   + (1-x) \Big( 6 - 2\ln(x) \Big)
\bigg].
\end{split}
\end{equation}
As we see IR part ($I_0$ term) cancels between
Br and Bx diagrams.
The same phenomenon is also
true for the differential distributions,
see figures and analytical investigation of the $\alpha_i\to 0$
limit in ref.~\cite{Slawinska:2009gn},
which are based on the above results.
The above IR cancellations are vital for the stability
of the NLO MC weight, see tests of the prototype MC
in ref.~\cite{Jadach:2010ew}.
It should be stressed that it is not guaranteed%
\footnote{Such IR cancellations would not work
 for virtuality ordering in the LO MC, $s(k_1,k_2)=-(p-k_1-k_2)^2$.}
and here it is true thanks to a good choice of the
multigluon LO MC distributions
compatible with the soft (eikonal) limit already at the LO level,
see eq.~(\ref{eq:Br12K}).

%\newpage
\subsection{Gluon pair production diagram - Vg}
%%%%%%%%%%%%%%%%%%%%%%%%%%%%%%%%%%%%%%%%%%%%%%%%%%%
\label{sec:Vg_res}
Let us investigate now another important 2R NLO contribution from
the gluon pair production diagram Vg
of Fig.~\ref{subfig:Vg}.
We shall calculate its contribution
to inclusive kernels focusing 
on possible differences between MC implementation and CFP scheme.
As we shall see, the 2R gluon distribution from Vg diagram
features very different singularity pattern than
Br+Bx diagrams discussed previously --
it has an internal collinear singularity for the
mass of produced pair going to zero, $\ba^2\to 0$,
that is located in $a_1\to a_2$ (instead of $a_1\to 0$).
Cancellation of this internal singularity happens
without an intervention of $(1-\Pbbm)$ operator,
simply by adding a virtual diagram (gluon selfenergy).
The remaining residual double logarithm
(or double pole in $\epsilon$) singularity is related
to the running coupling constant.

Mastering the additional soft-gluon sudakovian IR singularities
$\alpha_i\to0$
will be again very important for stability of the NLO MC weight.
A complete discussion of the soft gluon limit will require
introducing the interference diagrams Yg and Bx
of Fig.~\ref{subfig:Yg} and Fig.~\ref{subfig:Bx},
hence it will be deferred until the next section.

From the Feynman diagram we obtain:
\begin{equation}
\label{eq:WVg0}
\begin{split}
&\tilde{W}^{Vg} 
= \frac{4C_AC_F}{(1-x)^2}\frac{1}{(2\pi)^{4\epsilon}}
     \left(\frac{\alpha_S}{2\pi^2}\right)^2
     \frac{a_1^2a_2^2}{\tilde{q}^4(a_1,a_2)}
\\&~~~~~~~~\times
\bigg[ T_0^{Vg} 
     + T_{+}^{Vg}(\epsilon)\frac{\ba_1^2+\ba_2^2}{\ba^2}
     + T_{-}^{Vg}\frac{\ba_1^2-\ba_2^2}{\ba^2}
     + T_3^{Vg}(\epsilon)\frac{(\ba_1^2-\ba_2^2)^2}{\ba^4}
\bigg],
\\&
T_0^{Vg} = -2\alpha_1\alpha_2 +4(1-x)
   -(1-x)(2-x+x^2)
         \left(\frac{1}{\alpha_1}+\frac{1}{\alpha_2}\right),
\quad
T_3^{Vg}(\epsilon) = 2x(1+\epsilon),
\\&
T_{+}^{Vg}(\epsilon)=
(1-x)^2
 \bigg[
  (1+x^2) \Big( \frac{1}{\alpha_1^2}+\frac{1}{\alpha_2^2} \Big)
 +1 \bigg]
+\epsilon
 \bigg[
  (1-x)^4 \Big( \frac{1}{\alpha_1^2}+\frac{1}{\alpha_2^2} \Big)
 +(1-x)^2
 \bigg],
\\&
T_{-}^{Vg}=
(\alpha_1-\alpha_2)\bigg[
 (1+x)
-(2-x+x^2)     \frac{1-x}{\alpha_1\alpha_2}
+(1+x^2)\frac{(1-x)^3}{\alpha_1^2\alpha_2^2}
\bigg].
\end{split}
\end{equation}
% where $C=C_AC_F$.
In the above we have kept
only those terms ${\cal O}(\epsilon^1)$ from the $\gamma$-trace
which lead to $\epsilon\frac{1}{\epsilon^2}$
poles, because of the extra $\frac{1}{\epsilon}$ from
an internal $\frac{1}{a^2}$ gluon mass singularity.
Using
$
\frac{(\ba_1^2-\ba_2^2)^2}{\ba^4}
=\frac{\ba_1^2+\ba_2^2}{\ba^2}
+\frac{[(\ba_1+\ba_2)\cdot \ba ]^2-(\ba_1^2+\ba_2^2)\ba^2}{\ba^4}
$
the most singular term in eq.~(\ref{eq:WVg0})
is isolated even more clearly:
\begin{equation}
\label{eq:WVg}
\begin{split}
&\tilde{W}^{Vg} 
= \frac{4C_AC_F}{(1-x)^2}\frac{1}{(2\pi)^{4\epsilon}}
     \left(\frac{\alpha_S}{2\pi^2}\right)^2
     \frac{a_1^2a_2^2}{\tilde{q}^4(a_1,a_2)}
\bigg[ T_0^{Vg} 
  +T_{2+}^{Vg}(\epsilon) \; \frac{\ba_1^2+\ba_2^2}{\ba^2}
\\&~~~~~~~~~~~~~~~~~~~~~~~~~~~~~~~~
  +T_{-}^{Vg}\frac{\ba_1^2-\ba_2^2}{\ba^2}
  +T_3^{Vg}(0)\frac{[(\ba_1+\ba_2)\cdot \ba ]^2-(\ba_1^2+\ba_2^2)\ba^2}{\ba^4}
\bigg].
\end{split}
\end{equation}
The term proportional to 
$T_{2+}^{Vg}(\epsilon)= T_{+}^{Vg}(\epsilon)+T_3^{Vg}(\epsilon)$
is the only one contributing the double pole $\frac{1}{\epsilon^2}$
and is explicitly
proportional to the product of two LO kernels:
\begin{equation}
\begin{split}
&T_{2+}^{Vg}(\epsilon)
=\frac{(1-x)^3}{\alpha_1\alpha_2}\;\bigg\{
 \frac{1+x^2}{1-x}
\Big( \frac{\alpha_2}{\alpha_1}+\frac{\alpha_1}{\alpha_2}
     +\frac{\alpha_1\alpha_2}{(1-x)^2}
\Big)\\
&~~~~~~~~~
+\epsilon \Big[\frac{1+x^2}{1-x}\frac{\alpha_1\alpha_2}{(1-x)^2}
+(1-x)\Big(\frac{\alpha_2}{\alpha_1}+\frac{\alpha_1}{\alpha_2}\Big)
\Big]\bigg\}
=\frac{(1-x)^3}{\alpha_1\alpha_2}
 P^{[0]}_{qq}(x,\epsilon) P^{[0]}_{gg}(z) + 2x\epsilon,
\end{split}
\end{equation}
where $z=\alpha_1/(\alpha_1+\alpha_2)$ and
$ P^{(0)}_{gg}(z) = \frac{z}{1-z}+\frac{1-z}{z} +z(1-z)$.

What enters into the 2R part of the NLO correction in the MC,
see refs.~\cite{Jadach:2010aa,Jadach:2010ew},
is not the above divergent $\tilde{W}^{Vg}$,
but rather the non-divergent difference 
with the following ``soft collinear counterterm'' (SCC)
representing the distribution used in the LO MC:
% % \begin{equation}
% % \label{eq:VgK}
% % \tilde{K}^{Vg}_{MC} =
% %   \frac{C_F\alpha_s}{\pi^2}\frac{C_A\alpha_s}{\pi^2}\;
% %   \frac{1}{\mu^{4\epsilon}}
% %   \frac{2 a_1^2 a_2^2 }{ a_{\max}^2 \ba^2}
% % \frac{\alpha_1\alpha_2}{(1-x)}\;
% % P^{[0]}_{qq}(x,\epsilon) P^{[0]}_{gg}(z),
% % \end{equation}
% \begin{equation}
% \label{eq:VgK}
% \tilde{K}^{Vg}_{MC} =
%   \frac{C_F\alpha_s}{\pi^2}\frac{C_A\alpha_s}{\pi^2}\;
%   \frac{2 a_1^2 a_2^2 }{ a_{\max}^2 \ba^2}
% \frac{\alpha_1^2\alpha_2^2}{(1-x)^4}\;
% T_{2+}^{Vg}(0),
% \end{equation}
\begin{equation}
\label{eq:VgK}
\tilde{K}^{Vg} =
  \left(\frac{\alpha_S}{2\pi^2}\right)^2 4C_AC_F\;
  \frac{2 a_1^2 a_2^2 }{ a_{\max}^2 \ba^2}
\frac{\alpha_1^2\alpha_2^2}{(1-x)^4}\;
T_{2+}^{Vg}(0),
\end{equation}
where $a_{\max}=\max(a_1,a_2)$.
It is the result of a slight simplification
of the term $\sim T_{2+}^{Vg}$ in eq.~(\ref{eq:WVg}).
Generally, such a SCC is not unique, 
but complete NLO corrections to observables are insensitive
to its choice.
What is highly sensitive, however,
is the dispersion (and positiveness!) of the MC
weight implementing NLO correction.
The above choice is unique in this sense,
that it encapsulates not only collinear
singularity from $\ba^2\to 0$,
but also all soft gluon singularities $\alpha_i\to 0$
for all three diagrams Vg+Yg+Bx, see next section.
This property ensures good behavior of the NLO MC weight.
Moreover, the factor
$\frac{P^{[0]}_{gg}(z)}{\ba^2}$
in eq.~(\ref{eq:VgK})
can be iterated
into a separate final state LO sub-ladder
for the gluon emitted from the primary initial state ladder,
see refs.~\cite{Jadach:2010aa,Jadach:2010ew}.

As we are working on the exclusive level we are technically similar to the
techniques of hard process subtractions of~ref.~\cite{Catani:1996vz}
(dipole subtraction) or ref.~\cite{GehrmannDeRidder:2005cm} (antenna subtraction),%
\footnote{The dipole/antenna subtractions works for at least two ladders
(they are dealing with hard process), whereas our method works within
one ladder. Furthermore, we pay special attention to the fact that our
counterterms can be iterated in the MC simulation.}
but not to the subtractions used in the inclusive calculations of NNLO
evolution kernels of refs.~\cite{Moch:2004pa,Vogt:2004mw}.

In view of the above discussion, it is useful to know analytically,
for numerical cross-check of the MC code and for discussing
complete NLO corrections in the ladder MC,
the following subtracted Vg contribution to the NLO PDF%
\footnote{
  In the above $1/\alpha_i$ are regularized by a small
  parameter $\delta$ as in CFP, however,
  this is not necessary once diagrams Yg and Bx are added,
  see next section.}
calculated in $n=4$ (as usually including BE factor)
\begin{equation}
\label{eq:PVgSub}
\begin{split}
&
G^{Vg}_{sub}(Q,x)=
\frac{1}{2!} \int d\Phi_4(k_1) d\Phi_4(k_2)
\Big[ \tilde{W}^{Vg} - \tilde{K}^{Vg} \Big]
  \delta_{1-x=\alpha_1+\alpha_2}
  \theta_{Q>a_{max}>q_0}
\\&~~~~~~~~~~~
=\ln\frac{Q}{q_0}\; \Peu^{Vg}_{sub}(x),
\\&
% \Peu^{Vg}_{sub}(x)=
% \frac{2C_F \alpha_S}{\pi}\frac{2C_A \alpha_S}{\pi}\;
% \frac{1}{8}
% \Bigg\{
%     \frac{1}{3} \frac{x}{1-x}
%    +\frac{1+x^2}{1-x} \bigg[
%          \frac{\pi^2}{3}
%         -\frac{17}{9}
%         -2 I_1 -2I_0
% \\&~~~~
%         +2I_0 \ln\frac{1-x}{x}
%         +\frac{11}{6}\ln(x)
%         -2 \ln(1-x)
%         +\ln^2(1-x)
%         -2\ln(x)\ln(1-x)
%        \bigg]
% %    +\frac{1+x^2}{1-x} 4\bigg[
% %           I_0 +\ln (1-x)  -\frac{11}{12}
% %      \bigg] \ln\frac{1}{\varkappa}
% \Bigg\},
\Peu^{Vg}_{sub}(x)=
\left(\frac{\alpha_S}{2\pi}\right)^2 \Big(\frac{1}{2}C_AC_F\Big)
\Bigg\{
    \frac{4}{3} \frac{x}{1-x}
   +\frac{1+x^2}{1-x} \bigg[
        -8 I_1 -8I_0
        +8I_0 \ln\Big(\frac{1-x}{x}\Big)
\\&~~~~
        +\frac{22}{3}\ln(x)
        -8 \ln(1-x)
        +4\ln^2(1-x)
        -8\ln(x)\ln(1-x)
         \frac{4\pi^2}{3}
        -\frac{68}{9}
       \bigg]
\Bigg\}.
\end{split}
\end{equation}

Generally, the presence of the SCC subtraction is a natural
element in any MC scheme with soft gluon (photon)
resummation, either
to the hard process or to the ladders,
in order to eliminate possible double counting of the singular term.
However, the use of subtractions can also simplify non-MC calculations,
like analytical integration of the Vg diagram over the 2R phase space
in $n=4+2\epsilon$ dimensions in the CFP scheme,
before combining it with the gluon self-energy virtual diagram.
Let us comment more on that, venturing a little bit in the area
of combining 2R and 1R1V contributions
(complete discussion is beyond the scope of this work).
In such a case, it is useful to split 2R gluon phase space
into $a>\kappa a_{\max}$ and $a<\kappa a_{\max}$, $\kappa\ll 1$,
schematically
\[
\Gamma^{Vg} = \Gamma^{Vg}_{a>\kappa a_{\max}} +\Gamma^{Vg}_{a<\kappa a_{\max}}
\]
and split the Vg contribution into a subtracted one and the SCC.
The subtracted part in the decomposition
\[
\Gamma^{Vg} 
  = (\Gamma^{Vg}-\Gamma^{CT})+\Gamma^{CT}_{a>\kappa a_{\max}}
  + \Gamma^{CT}_{a<\kappa a_{\max}}
\]
can be evaluated in $n=4$, all over
the phase space.
On the other hand, the SCC part $\Gamma^{CT}$
is evaluated analytically separately
in the ``resolved'' part $a>\kappa a_{\max}$ in $n=4$,
and separately 
in the $a<\kappa a_{\max}$ part in $n=4+2\epsilon$.
This allows to profit from adjusting phase space parametrization
to specific complications of the integrand in each part!
Finally, one combines all three parts into a formula
for the 2R integrated Vg in $n=4+2\epsilon$.
The parameter $\kappa$, and even the dependence on
the particular choice of SCC drops out in the final result.
The other immediate profit from the above methodology is that
one may combine 2R from $a<\kappa a_{\max}$ with
the 1R1V contribution (gluon self-energy) such
that the Sudakovian part of the $\frac{1}{\epsilon^2}$ pole gets 
cancelled.%
\footnote{
  The remaining uncancelled part
  $\sim\frac{C_A\alpha_s}{\pi} \frac{11}{12}\frac{1}{\epsilon^2}$
  is related to the running coupling constant.}

This kind of calculation for
Vg in $n=4+2\epsilon$ is presented in the Appendix
using a slightly different choice (for historical reasons)
of the SCC:
\begin{equation}
\label{eq:WVgCT}
\begin{split}
&\tilde{W}^{Vg}_{CT} = 
  \frac{4C_AC_F}{(1-x)^2}
  \frac{1}{(2\pi)^{4\epsilon}}
  \left(\frac{\alpha_S}{2\pi^2}\right)^2
  \frac{a_1^2a_2^2}{\tilde{q}^4(a_{\max},a_{\max})}
  \bigg[ 
     T^{Vg}_{+}(\epsilon) \frac{2 a_{\max}^2}{\ba^2}
    +T_3^{Vg}(\epsilon)   \frac{(\ba_1^2-\ba_2^2)^2}{\ba^4} 
  \bigg].
\end{split}
\end{equation}
Switching from one kind of SCC
in the integration to another is relatively simple,
see Appendix.

Last but not least, let us discuss the differences
between the MC scheme and the CFP scheme for the Vg diagram.
In the previous case of Br diagram we have seen
that the basic mechanism of producing differences
between two schemes is the action of the $1-\Pbbm$
operator.
Since this operator is absent for Vg, one generally expects
no differences between the two schemes.
In particular any effect of
the terms proportional to $\epsilon$
from $\gamma$-traces will land in the $\sim\delta(\ba^2)$ part,
where soft 2R and 1R1V are combined together
in the same way in both schemes.

The only subtle point is the term
$b_0 \frac{1}{\epsilon^2}$ left over
from adding 2R soft and 1R1V contributions.
It comes from integrating over the $\ln\frac{q}{\mu_R}$ term
in the gluon self-energy,
and in the MC scheme it builds up an $\alpha_s$ dependence
for some kinematic variable.
Which variable? Changing from one choice of $q$ to another
may generate an extra NLO term in the kernel (in MC scheme).
Our preliminary study shows that taking transverse
momentum as $q$ is compatible with CFP,
that is the Vg diagram contribution
is then identical in MC and CFP.
The contribution from the diagrams Yg and Bx
discussed in the next section,
will contribute the same way in both schemes,
due to the lack of any internal collinear singularities.

%\newpage
\subsection{Gluon interference diagram - Yg}

%%%%%%%%%%%%%%%%%%%%%%%%%%%%%%%%%%%%%%%%%%%%%%%%%%%%%%%%%
The Monte Carlo distribution for the gluon interference diagram
of Fig.~\ref{subfig:Yg} (Yg) is given by:
\begin{equation}
\begin{split}
\label{eq:WYg}
&\tilde{W}^{Yg}(k_1,k_2) = \tilde{W}^{Yg1}(k_1,k_2) + \tilde{W}^{Yg1}(k_2,k_1),\\
&\tilde{W}^{Yg1}(k_1,k_2) = \frac{2}{1-x}\Big(\frac{1}{2}C_AC_F\Big)\left(\frac{\alpha_S}{2\pi^2}\right)^2
    \frac{\ba_1^2\ba_2^2}{\tilde{q}^4(a_1,a_2)}\\
&~~~~~~~~~~\times
      \bigg[T_0^{Yg} + T_1^{Yg}\frac{\ba_1\cdot\ba_2}{\ba_1^2}
    + T_3^{Yg}\frac{\ba_2^2-\ba_1^2}{\ba^2}
    + T_4^{Yg}\frac{\ba_2^2 (\ba_1\cdot\ba_2)-\ba_1^4}{\ba_1^2\ba^2}\bigg],
\end{split}
\end{equation}
where: %$C=\frac{1}{2}C_AC_F$ and:
\begin{equation}
\begin{split}
&T_0^{Yg} =  -2\alpha_1^2 + (5-4x)\alpha_1 + 2(x^2+2x-2)
     + \frac{4x^3-2x^2+3x-1}{\alpha_2} + \frac{x^2-x+2}{\alpha_1},\\
&T_1^{Yg} = 2\alpha_1^2 - 2(3-x)\alpha_1 + 2(3-x)
     + \frac{2(x^3-x^2+2x-2)}{\alpha_1},\\
% &T_2^{Yg} = 2(1-x) - \alpha_1 - \frac{x^2-x+2}{\alpha_1}
%      - \frac{2x^2+x+1}{\alpha_2},\\
&T_3^{Yg} = -4(2-x) + 3\alpha_1 + \frac{3x^2-5x+8}{\alpha_1}
     + \frac{2x^2+x+1}{\alpha_2} + \frac{4 (x^3-x^2+x-1)}{\alpha_1^2},\\
&T_4^{Yg} = 2(3-x) - 2\alpha_1 - \frac{2(x^2-2x+3)}{\alpha_1}
     - \frac{4(x^3-x^2+x-1)}{\alpha_1^2}.
\end{split}
\end{equation}
The expression for the contribution of the Yg diagram to the NLO
kernel is equal to (we include both factors $1/2!$ from BE
and 2 due to interference):
\begin{equation}
\begin{split}
&
G^{Yg}(Q,x)=
\int d\Phi_4(k_1) d\Phi_4(k_2)
\tilde{W}^{Yg}(k_1,k_2)\,
  \delta_{1-x=\alpha_1+\alpha_2}\,
  \theta_{\max\{a_1,a_2\}<a_{\max}}.
\end{split}
\end{equation}
% \begin{equation}
% \begin{split}
% &G^{Yg1}(Q,x) 
% = \int \frac{d\alpha_1}{\alpha_1}
%   \frac{d\alpha_2}{\alpha_2}\delta(1-x-\alpha_1-\alpha_2)
%   \int_0^{\infty}\frac{da_1}{a_1} \int_0^{\infty}\frac{da_2}{a_2}
%   \int_0^{2\pi} \frac{d\phi}{2\pi}
%   \;\theta_{\max\{a_1,a_2\}<a_{\max}}\\
% &\;\;\;\times \frac{2C}{1-x}\left(\frac{\alpha_S}{2\pi}\right)^2
%     \frac{\ba_1^2\ba_2^2}{\tilde{q}^4(a_1,a_2)}
%     \bigg[T_0^{Yg} + T_1^{Yg}\frac{\ba_1\cdot\ba_2}{\ba_1^2}
%     + T_3^{Yg}\frac{\ba_2^2-\ba_1^2}{\ba^2}
%     + T_4^{Yg}\frac{\ba_2^2 (\ba_1\cdot\ba_2)-\ba_1^4}{\ba_1^2\ba^2}\bigg].
% \end{split}
% \end{equation}
The integrand only has the familiar scale singularity which can
be extraced in standard way as a logarithm.
Then the integrals over transverse components take form:
\begin{equation}
\begin{split}
&G^{Yg}(Q,x) 
= \ln\frac{Q}{q_0} \frac{4}{1-x}\Big(\frac{1}{2}C_AC_F\Big)
   \left(\frac{\alpha_S}{2\pi}\right)^2
   \int \frac{d\alpha_1}{\alpha_1} \frac{d\alpha_2}{\alpha_2}
   \delta(1-x-\alpha_1-\alpha_2)\\
&\;\;\;\times \bigg\{ T_0^{Yg}\frac{\alpha_1\alpha_2}{2x}
   - T_1^{Yg}\frac{\alpha_1^2\alpha_2}{2x(1-\alpha_1)}
   + T_3^{Yg}\frac{\alpha_1\alpha_2(\alpha_1-\alpha_2)}{2x(1-x)}\\
&\;\;\; + T_4^{Yg}\bigg[-\frac{\alpha_1^2}{2}
   \ln\bigg(\frac{(1-x)(1-\alpha_1)}{x\alpha_1}\bigg)
   - \frac{\alpha_1\alpha_2(3\alpha_1^2-2\alpha_1+\alpha_2)}
   {2x(1-x)(1-\alpha_1)}\bigg] \bigg\}.
\end{split}
\end{equation}
After the final integration over $\alpha_i$ variables we obtain
\begin{equation}
\begin{split}
&G^{Yg}(Q,x) = \ln\frac{Q}{q_0} \,\Peu^{Yg}(x) ,\\
&\Peu^{Yg}(x) = \left(\frac{\alpha_S}{2\pi}\right)^2
     \left(\frac{1}{2}C_AC_F\right)
     \bigg\{ \frac{1+x^2}{1-x}\bigg[ 8I_1 + 16I_0
     - 8I_0\ln\Big(\frac{1-x}{x}\Big)
     - 2\ln^2(x)
\\&~~~
     - 4\ln^2(1-x) + 4\ln(x)\ln(1-x)
     - 3\ln(x) + 16\ln(1-x)
     - 9 - 4\Li_2(x) \bigg]
\\&~~~~
     + (1+x)\ln(x) + 3(1-x)
     + \frac{2}{1-x} \bigg\}.
\end{split}
\label{eq:P_Yg}
\end{equation}

Gluon interference diagram Yg features both double and single
logarithmic IR divergences ($I_1$ and $I_0$). 
Of course, they must cancel when all diagrams are added.
Cancellations occur for 2R diagrams not only on the
inclusive integrated level but already on the exclusive
unintegrated level, see~\cite{Slawinska:2009gn}.
We can see them explicitly by adding inclusive kernel contributions
for gluon interference diagram Yg~\eqref{eq:P_Yg}, subtracted
gluon pair production diagram Vg~\eqref{eq:PVgSub} and
part of gluonstrahlung interference diagram Bx~\eqref{eq:BxP}
proportional to $\frac{1}{2}C_AC_F$ colour factor:
\begin{equation}
\begin{split}
&\Peu^{Yg}(x) + \Peu^{Vg}_{sub}(x) + \Peu^{Bx}(x)
   = \left(\frac{\alpha_S}{2\pi}\right)^2
     \left(\frac{1}{2}C_AC_F\right)
\\&~~~~~
     \times\bigg\{ \frac{1+x^2}{1-x}\bigg[ 
     - 4\ln(x)\ln(1-x) + \frac{13}{3}\ln(x)
     + \frac{4\pi^2}{3} 
     - \frac{149}{9} - 4\Li_2(x) \bigg]
\\&~~~~~
     - 3(1+x)\ln(x) + 3(1-x)
     + \frac{2}{1-x} + \frac{4}{3}\frac{x}{1-x} \bigg\}.
\end{split}
\label{eq:P_Yg+Vgsub+Bx}
\end{equation}
The above expression for the sum of 2R contribution with colour
factor equal $\frac{1}{2}C_AC_F$ is free from double and single
logarithmic IR divergences ($I_1$ and $I_0$), as expected.

%%%%%%%%%%%%%%%%%%%%%%%%%%%%%5
%%%%%%%%%%%%%%%%%%%%%%%%%%%%%5
\section{Other non-singlet diagrams}
\label{sec:unres}
%%%%%%%%%%%%%%%%%%%%%%%%%%%%%5

The contributions to NLO kernels from diagrams displayed in
Fig.~\ref{subfig:Vf}, Fig.~\ref{subfig:Yf} and Fig.~\ref{subfig:Xf}
will be presented in the following.
They will be referred to as Vf, Yf and Xf diagrams respectively.
As discused in the introduction,
before all singlet class diagrams are included,
the  contributions from the interference diagrams Xf and Yf
should enter the MC code in the inclusive form.

%=================================================
\subsection{ Interference diagram Xf}

The crossed-ladder diagram Xf
contributes the following distribution
\begin{equation}
\begin{split}
\label{eq:WXf}
\tilde W^{Xf}(k_1, k_2) &= \Big(C_F^2-\frac{1}{2}C_AC_F\Big)
\left(\frac{\alpha_S}{2\pi^2}\right)^2
\\&\times
\frac{a_1^2 a_2^2}{\tilde{q}^4(a_1,a_2)}
\left( 
T_0^{Xf} 
+ T_{12}^{Xf}\cos^2\phi 
+ T_1^{Xf}\frac{a_2}{a_1}  \cos\phi
+ T_2^{Xf}\frac{a_1}{a_2}  \cos\phi
\right)
\end{split}
\end{equation}
to the quark--antiquark kernel, where:
\begin{equation}
\begin{split}
T_0^{Xf}= -\frac{(\alf1 - \alf2)^2 x}{(\alf1-1)(\alf2-1)},
\quad &
T_{12}^{Xf} = 2 (x^2 + 1)\frac{\alf1 \alf2}{(\alf1 - 1)(\alf2 -1)},
\\
T_1^{Xf} 
=  (x+1)\frac{\alf2(\alf1^2+\alf1 \alf2-\alf1+\alf2)}{(\alf1-1)(\alf2-1)},
\quad &
T_2^{Xf} 
= (x+1)\frac{\alf1(\alf2^2+\alf1 \alf2 + \alf1 - \alf2)}{(\alf1-1)(\alf2-1)}.
\end{split}
\end{equation}
% and $C = C_F^2 - C_FC_A/2$.
The integrated distribution is equal to:
\begin{equation}
\begin{split}
\Peu^{Xf}_{q\bar{q}}(x) =& -\left( \frac{\alpha_S}{2\pi}\right)^2
 \left(C_F^2 - \frac{1}{2}C_AC_F\right)
 \bigg[ 4(1+x)\ln(x) + 8(1-x) \\
& +\frac{1+x^2}{1+x}\bigg(
2\ln^2(x) - 4\ln(1+x)\ln(x) 
+ 4\Li_2\bigg( \frac{x}{1+x}\bigg)- 4\Li_2\bigg( \frac{1}{1+x}\bigg)
\bigg)
\bigg]
\end{split}
\end{equation}
and the contribution to the NLO kernel equals:
\begin{equation}
G^{Xf}(Q,x)=\ln\frac{Q}{q_0}\Peu^{Xf}_{q\bar{q}}(x).
\end{equation}

% Where the integral used in (4.55) of \cite{Curci:1980uw} equals:
% \begin{equation}
% \int_{\frac{x}{x+1}}^{\frac{1}{x+1}}\log\bigg( \frac{1-z}{z}\bigg)\frac{d z}{z}
% =\frac{1}{2}\log^2(x) - \log(x+1)\log(x) 
% + Li_2\bigg( \frac{x}{x+1}\bigg)- Li_2\bigg( \frac{1}{x+1}\bigg)
% \end{equation}

The above agrees with \cite{Curci:1980uw} up to the $-$ 
sign which is a matter of convention.

%=================================================
\subsection{Fermion interference diagram - Yf}
The differential distribution for the fermion interference diagram
Yf of Fig.~\ref{subfig:Yf} reads:
\begin{equation}
\label{eq:WYf}
\begin{split}
&\tilde{W}^{Yf} = \frac{2}{1-x}\Big(C_F^2-\frac{1}{2}C_AC_F\Big)
    \left(\frac{\alpha_S}{2\pi^2}\right)^2
    \frac{\ba_1^2\ba_2^2}{\tilde{q}^4(a_1,a_2)}\\
&\;\;\;\times \bigg[ T_0^{Yf}
    + T_1^{Yf}\frac{\ba_1\cdot\ba_2}{\ba_1^2}
    + T_3^{Yf}\frac{\ba_2^2-\ba_1^2}{\ba^2}
    + T_4^{Yf}\frac{\ba_2^2\,(\ba_1\cdot\ba_2)-\ba_1^4}{\ba_1^2\,\ba^2}
    \bigg],
\end{split}
\end{equation}
% The colour factor is equal to $C=C_F^2-\frac{1}{2}C_AC_F$ and
where:
\begin{equation}
\begin{split}
&T_0^{Yf} = \frac{1-4\alpha_1\alpha_2-\alpha_2^2
     +2\alpha_1\alpha_2^2}{1-\alpha_1},\\
&T_1^{Yf} = \frac{2\alpha_2\left(1-2\alpha_1\alpha_2
     -\alpha_2^2\right)}{1-\alpha_1},\\
% &T_2^{Yf} = \frac{\alpha_2^2+2\alpha_1-1}{1-\alpha_1},\\
&T_3^{Yf} = 
     \frac{2\alpha_2^3+3\alpha_1\alpha_2^2-4\alpha_2^2
     +4\alpha_1^2\alpha_2-4\alpha_1\alpha_2+2\alpha_2
     -2\alpha_1^2+\alpha_1}{(1-\alpha_1)\alpha_1},\\
&T_4^{Yf} =
     \frac{2\alpha_2\left(-2\alpha_1^2+2(1-\alpha_2)
     \alpha_1-\alpha_2^2+2\alpha_2-1\right)}
     {(1-\alpha_1)\alpha_1}.
\end{split}
\end{equation}
The contribution of the Yf interference diagram to the inclusive
kernel (bare PDF) is given by (including interference factor 2):
\begin{equation}
\begin{split}
&G^{Yf}(Q,x) = \ln\frac{Q}{q_0} \,\Peu^{Yf}(x),\\
&\Peu^{Yf}(x) = \left(\frac{\alpha_S}{2\pi}\right)^2 %tu jest 2 od interferencji
    \Big(C_F^2-\frac{1}{2}C_AC_F\Big)\\
&\;\;\;\;\;\;\;\;\;\;\;\;\times
    \bigg\{ \frac{1+x^2}{1-x}\bigg[ 2\ln^2(x) - 4\ln(x)\ln(1-x)
  + 3\ln(x) + \frac{2\pi^2}{3} - 4\Li_2(x)\bigg]\\
&\;\;\;\;\;\;\;\;\;\;\;\; + 15(1-x)
  + (1+x)\left(1+7\ln(x)\right) \bigg\}.
\end{split}
\end{equation}
This result agrees with that of ref.~\cite{Heinrich:1997kv},%
\footnote{The authors use a different regularization technique.
In case of the Yf diagram, however, both methods give the same
result, as explained in \cite{Heinrich:1997kv}.}
the difference in sign can be attributed to a different definition
of space dimension ($n=4-2\epsilon$, $\epsilon$ being negative as opposed to CFP).

%=================================================
\subsection{Fermion pair production diagram - Vf}
\label{sec:Vf_res}
The fermion pair production diagram Vf of Fig.~\ref{subfig:Vf}
features an internal collinear singularity when the mass of the
produced pair goes to zero. The kinematical structure of the Vf graph
is quite similar to that of the gluon pair production diagram Vg.
The $n=4+2\epsilon$ dimensional distribution for the Vf diagram reads:
\begin{equation}
\begin{split}
\label{eq:WVf}
&\tilde{W}^{Vf} 
= \frac{4}{(1-x)^2}\Big(\frac{1}{2}N_FC_F\Big)\frac{1}{(2\pi)^{4\epsilon}}
     \left(\frac{\alpha_S}{2\pi^2}\right)^2
     \frac{a_1^2a_2^2}{\tilde{q}^4(a_1,a_2)}
\\&~~~~~~~~~~~\times
\bigg[ T_0^{Vf} 
     + T_1^{Vf}(\epsilon)\frac{\ba_1^2}{\ba^2}
     + T_2^{Vf}(\epsilon)\frac{\ba_2^2}{\ba^2}
     + T_3^{Vf}\frac{(\ba_1^2-\ba_2^2)^2}{\ba^4}
\bigg],
\end{split}
\end{equation}
where: %the color factor $C=\frac{1}{2}N_FC_F$ and
\begin{equation}
\begin{split}
&T_0^{Vf} = -2\alpha_1^2+2(1-x)\alpha_1-(1-x)(1+x)%-\epsilon(1-x)^2
,\\
&T_1^{Vf}(\epsilon) = -2(1+x)\alpha_1+2x(1-x)+\frac{(1-x)(1+x^2)}{\alpha_2}
     + \epsilon\frac{(1-x)^3}{\alpha_2},\\
&T_2^{Vf}(\epsilon) = 2(1+x)\alpha_1-2(1-x)+\frac{(1-x)(1+x^2)}{\alpha_1}
     + \epsilon\frac{(1-x)^3}{\alpha_1},\\
&T_3^{Vf} = -2x.
\end{split}
\end{equation}
The contribution of this diagram integrated in
$n=4+2\epsilon$ dimensions reads:
\begin{equation}
\begin{split}
&\Gamma^{Vf} = \frac{\frac{1}{2}N_FC_F}{2\epsilon}\left(\frac{\alpha_S}{2\pi}\right)^2
    \frac{1+x^2}{1-x}\bigg[ \frac{2}{3}\bigg(\frac{1}{\epsilon}
  + 2\ln\left(\frac{Q^2}{4\pi\mu^2}\right) + 2\gamma\bigg)
  + \frac{8}{3}\ln(1-x) - \frac{2}{3}\ln(x) - \frac{10}{9}\bigg].
\end{split}
\end{equation}
For exclusive modeling of this diagram
we need to deal with its internal collinear singularity in
a similar way as for the gluon pair production diagram Vg.
The following counterterm
\begin{equation}
\begin{split}
\tilde{W}^{Vf}_{CT} &= 
  \frac{4}{(1-x)^2}\Big(\frac{1}{2}N_FC_F\Big)
  \frac{1}{(2\pi)^{4\epsilon}}
  \left(\frac{\alpha_S}{2\pi^2}\right)^2
\\&\times
  \frac{a_1^2a_2^2}{\tilde{q}^4(a_{\max},a_{\max})}
  \bigg[ 
     (T^{Vf}_1(\epsilon)+T^{Vf}_2(\epsilon)) \frac{a_{\max}^2}{\ba^2}
    +T_3^{Vf} \frac{(\ba_1^2-\ba_2^2)^2}{\ba^4} 
  \bigg]
\end{split}
\end{equation}
can be used both for the MC purpose in $n=4$ and for combining
real and virtual contributions.
In the latter case, we decompose the Vf contribution
into a part entering Monte Carlo
$(\Gamma^{Vf}-\Gamma^{Vf\,CT})+\Gamma^{Vf\,CT}_{a>\kappa a_{\max}}$
and an unresolved part $\Gamma^{Vf\,CT}_{a<\kappa a_{\max}}$ required
for the cancellation of double poles from the virtual contributions.
For completeness we give the subtracted contribution of the Vf diagram
to the NLO PDF:
\begin{equation}
\label{eq:PVfSub}
\begin{split}
&
G^{Vf}_{sub}(Q,x)=
\int d\Phi_4(k_1) d\Phi_4(k_2)
\Big[ \tilde{W}^{Vf} - \tilde{W}^{Vf}_{CT} \Big]
  \delta_{1-x=\alpha_1+\alpha_2}
  \theta_{Q>a_{max}>q_0}
=\ln\frac{Q}{q_0}\; \Peu^{Vf}_{sub}(x),
\\&
\Peu^{Vf}_{sub}(x)=
\left(\frac{\alpha_S}{2\pi}\right)^2 \left(\frac{1}{2}N_FC_F\right)
     \frac{1+x^2}{1-x} \bigg[
         \frac{14}{9} - \frac{4}{3}\ln(x)
       \bigg].
\end{split}
\end{equation}

%%%%%%%%%%%%%%%%%%%%%%%%%%%%%5
%%%%%%%%%%%%%%%%%%%%%%%%%%%%%5
\section{Summary and outlook}
%%%%%%%%%%%%%%%%%%%%%%%%%%%%%5
\label{sec:conclusios}

The main result of this work is a complete collection of
2-real parton (quark, gluon) differential
(unintegrated) distributions, which enter calculations
of the NLO DGLAP non-singlet
evolution kernels, in a form ready for the use
in the Monte Carlo implementation of the ladder
(also referred to as NLO parton shower MC).
The distributions are given in
% eqs.~(\ref{eq:BxW},\ref{eq:Br12W},\ref{eq:WVg0},\ref{eq:WYg},%
% \ref{eq:WXf},\ref{eq:WYf},\ref{eq:WVf}).
eqs.~\eqref{eq:BxW}, \eqref{eq:Br12W}, \eqref{eq:WVg0}, \eqref{eq:WYg},
\eqref{eq:WXf}, \eqref{eq:WYf}, \eqref{eq:WVf}.
These distributions in the fully differential form are not available
in the literature.
% This work is now being extended to the remaining singlet diagrams.

We also present the differential distribution 
of the collinear soft counterterms,
which are used to subtract
internal singularities for some diagrams.
These subtractions are also present in the MC weights.
The MC collinear soft counterterm distributions are defined in
% eqs.~(\ref{eq:Br12K},\ref{eq:VgK}).
eqs.~\eqref{eq:Br12K}, \eqref{eq:VgK}.

Furthermore, we present analytical integration results.
They are presented as the contributions to evolution kernels from
the same 2-real parton differential distributions listed above,
see for example eqs.~(\ref{eq:GMCBrX}) for all bremsstrahlung diagrams.
In case of diagrams with internal collinear singularities,
subtractions of the MC collinear counterterms is done.
For certain diagrams it was possible to compare
the integration with available published results
of refs.~\cite{Curci:1980uw,Heinrich:1997kv,Bassetto:1998uv}
and agreement was found.

The QCD evolution of the NLO ladder implemented 
in the MC is slightly different
from that of standard $\overline{MS}$, as defined and implemented in
Curci-Furmanski-Petronzio paper~\cite{Curci:1980uw}.
For instace,
the differences between MC and CFP schemes are discussed as far as
it is possible for the 2-real contributions.
They are typically present in the diagrams with internal,
collinear divergences, see for instance eq.~(\ref{eq:CFPvsMCBr}).
The complete discussion of this issue is beyond the scope of the present
work -- it will be completed when diagrams with 1 real and 1 virtual
corrections are added into the game in the forthcoming work.
Nevertheless, even incomplete results provide us important insight
into the differences between
NLO (integrated) kernels of MC and CFP $\overline{MS}$ schemes.%
\footnote{Luckily, all these differences are coming from small subset
of diagrams and are relatively simple.}
This analysis will also be a practical outcome of the entire project.

We did not explicitly show results of the numerical cross-checks of
the analytical results.
Let us only mention that
all analytical integration results in the paper were cross-checked
(up to 4-digits) by means of the MC numerical integration using {\tt FOAM}
program~\cite{foam:2002} within the {\tt MCdevelop}
system~\cite{Slawinska:2010jn}.

Summarizing, the present work marks an important step forward on the way
to the implementation of the complete NLO DGLAP ladder in the Monte Carlo form.

%%%%%%%%%%%%%%%%%%%%%%%%%%%%%%%%%%%%%%%%%%%%%%%%%%%%%%%%%%%%%%
%%%%%%%%%%%%%%%%%%%%%%%%%%%%%%%%%%%%%%%%%%%%%%%%%%%%%%%%%%%%%%
%%%%%%%%%%%%%%%%%%%%%%%%%%%%%%%%%%%%%%%%%%%%%%%%%%%%%%%%%%%%%%
\acknowledgments
This work is partly supported by 
the Polish Ministry of Science and Higher Education grants
No.\ 153/6.PR UE/2007/7 and N N202 128937 and by
the EU Framework Programme grant MRTN-CT-2006-035505
and by DOE grant DE-FG02-09ER41600.
One of the authors (S.J.) is grateful for partial support
and warm hospitality of TH Unit of CERN PH division,
and Physics Department of Baylor University,
while completing this work.

%%%%%%%%%%%%%%%%%%%%%%%%%%%%%%%
%%%%%%%%%%%%%%%%%%%%%%%%%%%%%%%
% \clearpage
%\vfill\newpage
%\noindent
%{\bf\Large APPENDIX}
\appendix
\numberwithin{equation}{section}
%%%%%%%%%%%%%%%%%%%%%%%%%%%%%%%

%\input appendix/appendix.tex
%=================================================
\section{Gluon pair production diagram - Vg}
\label{app:Vg}

Let us present details of the calculation of the evolution kernel
contribution for the Vg diagram.
First, we shall show the calculation for Vg subtracted
with the counterterm of eq.~(\ref{eq:WVgCT}) in $n=4$,
then we shall show how to switch to the MC counterterm
of eq.~(\ref{eq:VgK}) and finally we will
integrate Vg over the collinear part of the phase space,
$a<\kappa a_{\max}$, in $n=4+2\epsilon$ dimensions.

\subsection{Vg - subtracted}
%%%%%%%%%%%%%%%%%%%%%%%%%%%%%%%%%%%%%%%%%%%%%%%%%%%%%%%%%%%%
\label{app:Vg_subt}
The difference of Vg and the counterterm
of eq.~(\ref{eq:WVgCT}) has no collinear
divergence and in $n=4$ dimensions reads:
\begin{equation}
\begin{split}
&G^{Vg-CT}(Q,x)=
 \ln\frac{Q}{q_0}
 \frac{4C_AC_F}{(1-x)^2} 
 \left(\frac{\alpha_S}{2\pi}\right)^2
 \int \frac{d\alpha_1}{\alpha_1}\frac{d\alpha_2}{\alpha_2}
 \delta_{1-x=\alpha_1+\alpha_2}
 \int_0^{2\pi} \frac{d\phi}{2\pi}
\\&~~\times 
\bigg\{
     \int_0^1 \frac{dy_1}{y_1} \bigg[
     \frac{y_1^2}{\tilde{q}^4(y_1,1)}
     \bigg( T_0^{Vg} + \frac{T_1^{Vg}y_1^2+T_2^{Vg}}{\tilde{a}^2(y_1,1)}
     + T_3^{Vg}\frac{(y_1^2-1)^2}{\tilde{a}^4(y_1,1)} \bigg)
\\&~~~~~~~~~~~~~~~~~~~~~~~~~~~~~~~~~~~~~~~
     - \frac{y_1^2}{\tilde{q}^4(1,1)}
     \bigg( \frac{T_1^{Vg}+T_2^{Vg}}{\tilde{a}^2(y_1,1)}
     + T_3^{Vg}\frac{(y_1^2-1^2)^2}{\tilde{a}^4(y_1,1)} \bigg)\bigg]
\\
&~~~~~
  + \int_0^1 \frac{dy_2}{y_2} \bigg[
     \frac{y_2^2}{\tilde{q}^4(1,y_2)}
     \bigg( T_0^{Vg} + \frac{T_1^{Vg}+T_2^{Vg}y_2}{\tilde{a}^2(1,y_2)}
     + T_3^{Vg}\frac{(1-y_2^2)^2}{\tilde{a}^4(1,y_2)} \bigg)
\\&~~~~~~~~~~~~~~~~~~~~~~~~~~~~~~~~~~~~~~~
     - \frac{y_2^2}{\tilde{q}^4(1,1)}
     \bigg( \frac{T_1^{Vg}+T_2^{Vg}}{\tilde{a}^2(1,y_2)}
     + T_3^{Vg}\frac{(1-y_2^2)^2}{\tilde{a}^4(1,y_2)}
     \bigg) \bigg]\bigg\},
\end{split}
\end{equation}
where $\tilde{a}^2(y_1,y_2)=y_1^2+y_2^2-2y_1y_2\cos\phi_{12}$ is
a dimensionless function, $T^{Vg}_1= T^{Vg}_{2+}+T^{Vg}_{2-}$
and   $T^{Vg}_2= T^{Vg}_{2+}-T^{Vg}_{2-}$.
Performing integrations over the
transverse momenta variables $\phi$, $y_1$ and $y_2$ yields:
\begin{equation}
\begin{split}
G^{Vg-CT}&(Q,x)=
 \ln\frac{Q}{q_0}
 \frac{4C_AC_F}{(1-x)^2}
 \left(\frac{\alpha_S}{2\pi}\right)^2
 \int \frac{d\alpha_1}{\alpha_1}\frac{d\alpha_2}{\alpha_2}
 \delta_{1-x=\alpha_1+\alpha_2}
\\&~~\times
 \bigg\{ 
    T_0^{Vg}\frac{\alpha_1\alpha_2}{2x}
    + T_1^{Vg}\bigg[\frac{\alpha_1^2\alpha_2^2}{2(1-x)^2}
      \bigg( \ln\left(\frac{(1-x)^2}{x\alpha_1\alpha_2}\right)
      - 1\bigg) + \frac{\alpha_1\alpha_2^3}{2x(1-x)^2} \bigg]
\\&~~~~~~
  + T_2^{Vg}\bigg[\frac{\alpha_1^2\alpha_2^2}{2(1-x)^2}\bigg(
     \ln\left(\frac{(1-x)^2}{x\alpha_1\alpha_2}\right) - 1\bigg)
+ \frac{\alpha_1^3\alpha_2}{2x(1-x)^2} \bigg] 
\\&~~~~~~
  + T_3^{Vg} 
    \bigg[ \frac{\alpha_1\alpha_2(\alpha_1^2+\alpha_2^2)}
       {2(1-x)} + \frac{\alpha_1\alpha_2(\alpha_1^3+\alpha_1^2\alpha_2
       +\alpha_1\alpha_2^2-2\alpha_1\alpha_2+\alpha_2^3)}{2x(1-x)}
\\&~~~~~~~~~~~~~~~~~~~~~~~~~~~
  + \frac{\alpha_1\alpha_2(\alpha_1^2+\alpha_2^2)}{2(1-x)^2}
     + \frac{\alpha_1^2\alpha_2^2}{(1-x)^2}
    \ln\left(\frac{(1-x)^2}{x\alpha_1\alpha_2}\right) 
  \bigg]
\bigg\}.
\end{split}
\end{equation}
Finally the $\alpha_i$ integrations result is:%
\footnote{We always implicitly use {\em Principal Value} type
of regularization for $\alpha$ integrals.}
\begin{equation}
\label{eq:VgSubCT9}
\begin{split}
&G^{Vg-CT}(Q,x)=
 \ln\frac{Q}{q_0}
 C_AC_F
 \left(\frac{2\alpha_S}{\pi}\right)^2 
 \frac{1}{8}\bigg\{
     \frac{1+x^2}{1-x}\bigg[ -2I_1 - 2I_0 + 2I_0\ln(1-x)
    - 2I_0\ln(x)
\\&~~~~~~
 +  \ln^2(1-x) - 2\ln(x)\ln(1-x) - 2\ln(1-x)
    + \frac{11}{6}\ln(x) + \frac{\pi^2}{3} - \frac{25}{18}
    \bigg] - \frac{1}{2}(1-x) 
 \bigg\}.
\end{split}
\end{equation}

\subsection{Two counterterms for Vg}
%%%%%%%%%%%%%%%%%%%%%%%%%%%%%%%%%%%%%%%%%%%%%%%%%%%%%%%%%%%%

%%%%%%%%%%%%%%%%%%%%%%%%%%%%%%%%%%%%%%%%%%%%%%%%%%%%%%%%%%%
%%%%%%%%%%%%%%%%%%%%%%%%%%%%%%%%%%%%%%%%%%%%%%%%%%%%%%%%%%%
\begin{figure}[hhh]
  \begin{center}
    \includegraphics[width=5cm]{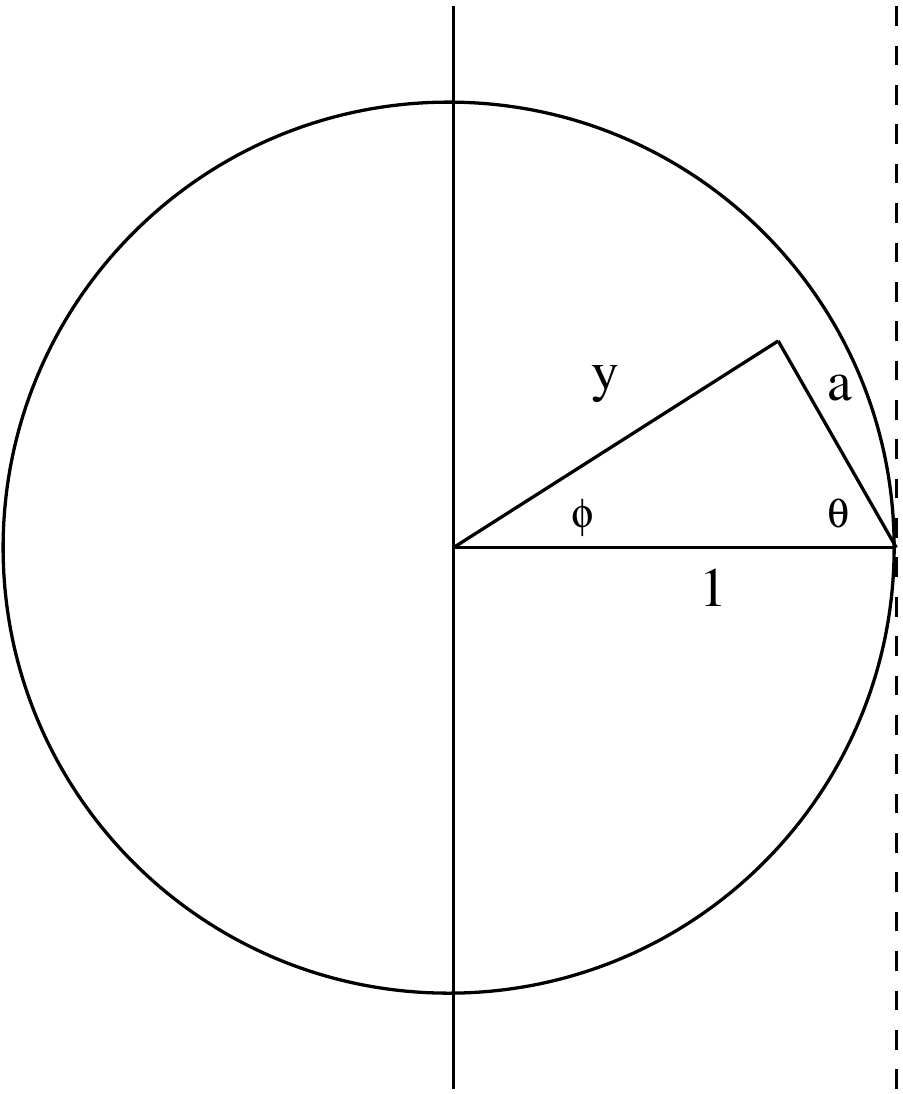}
  \end{center}
\caption{ The disc represents the region $|\ba_1|<|\ba_2|$.
The dimensionless ratio of $y=|\ba_1|/|\ba_2|$
and $\phi$, the relative angle between $\ba_1$ and $\ba_2$,
are shown.
Alternative variables ($\tilde{a}$, $\theta$) are also shown.}
\label{fig:a_theta}
\end{figure}
%%%%%%%%%%%%%%%%%%%%%%%%%%%%%%%%%%%%%%%%%%%%%%%%%%%%%%%%%%%
%%%%%%%%%%%%%%%%%%%%%%%%%%%%%%%%%%%%%%%%%%%%%%%%%%%%%%%%%%%

For various purposes it is useful to calculate the contribution
from both counterterms of 
eqs.~(\ref{eq:WVgCT}) and (\ref{eq:VgK}) 
in $n=4$ with the cutoff $|\ba|>\kappa a_{\max}$
($\tilde{a}>\kappa$).
Let us start with the counterterm of eq.~(\ref{eq:WVgCT}).
Introducing dimensionless variables
$y_i=a_i/a_{\max}$, $\tilde{a}=|\ba|/a_{\max}$
and performing integration
over the overall scale $a_{\max}$ we obtain:
\begin{equation}
\begin{split}
&G^{CT}_{\tilde{a}>\kappa} =
  \ln\frac{Q}{q_0}
  \frac{4C_AC_F}{(1-x)^2} 
  \left(\frac{\alpha_S}{2\pi}\right)^2
  \int \frac{d\alpha_1}{\alpha_1}\frac{d\alpha_2}{\alpha_2}
  \delta_{1-x=\alpha_1+\alpha_2}
  \int \frac{d\phi}{2\pi}\\
&~~~\times  
  \bigg[
     \int_0^1 dy_1
     \frac{y_1}{\tilde{q}^4(1,1)}
     \bigg( \frac{T^{Vg}_{+}}{\tilde{a}^2(y_1,1)}
     + T_3^{Vg}\frac{(y_1^2-1)^2}{\tilde{a}^4(y_1,1)} \bigg)\\
&~~~~~~ 
   + \int_0^1 dy_2
     \frac{y_2}{\tilde{q}^4(1,1)}
     \bigg( \frac{T^{Vg}_{+}}{\tilde{a}^2(1,y_2)}
     + T_3^{Vg}\frac{(1-y_2^2)^2}{\tilde{a}^4(1,y_2)}
     \bigg) 
  \bigg] \theta_{\tilde{a}>\kappa}.
\end{split}
\end{equation}
We need to remember that the $\kappa$ cutoff is 
infinitesimal and  all terms ${\cal O}(\kappa)$ are neglected.
Both $y_1$ and $y_2$ integrals are equal because
$\tilde{a}^2(y_1,y_2)$ is symmetric and 
$\tilde{q}^2(1,1)=\frac{(1-x)^2}{\alpha_1\alpha_2}$ 
does not depend on $y_i$.
\begin{equation}
\begin{split}
&G^{CT}_{\tilde{a}>\kappa} =
  \ln\frac{Q}{q_0}
  \frac{8C_AC_F}{(1-x)^4}
  \left(\frac{\alpha_S}{2\pi}\right)^2
  \int d\alpha_1d\alpha_2 \alpha_1\alpha_2
  \delta_{1-x=\alpha_1+\alpha_2}
\\&~~~~~~~~~\times
  \int_0^{2\pi} \frac{d\phi}{2\pi}
  \int dy_1 y_1
  \bigg( \frac{T^{Vg}_{+}}{\tilde{a}^2(y_1,1)}
     + T_3^{Vg}\frac{(y_1^2-1)^2}{\tilde{a}^4(y_1,1)} 
  \bigg) \theta_{\tilde{a}>\kappa}.
\end{split}
\end{equation}
The integration over $\alpha$ factorizes from the 
$y$ and $\phi$ integrals, hence
it is performed separately.
Moreover, because of the cutoff on $\tilde{a}^2$ it is convenient
to change variables.
Instead of $(y,\phi)$ variables we
use $(\tilde{a},\theta)$ depicted in Fig.~\ref{fig:a_theta}.
The jacobian for this transformation is equal to
$\tilde{a}/\sqrt{1+\tilde{a}^2-2\tilde{a}\cos\theta}$.
The two integrals above are calculated separately.
Firstly,
\begin{equation}
\begin{split}
&\frac{1}{2\pi}\int_0^{2\pi} d\phi \int_0^{1} dy
    \frac{y}{1+y^2-2y\cos\phi} \;\theta(\tilde{a}>\kappa)
  = \frac{1}{2\pi} \int_{\kappa}^2 \frac{d\tilde{a}}{\tilde{a}}
    \;2\int_0^{\arccos(\tilde{a}/2)} d\theta
\\&~~
  = \frac{1}{\pi} \int_{\kappa}^2 d\tilde{a}
    \frac{\arccos(\tilde{a}/2)}{\tilde{a}}
  = \frac{1}{2}\ln\frac{1}{\kappa} + {\cal O}(\kappa)
\end{split}
\end{equation}
and next
\begin{equation}
\begin{split}
&\frac{1}{2\pi}\int_0^{2\pi} d\phi \int_0^{1} dy
    \frac{y(1-y^2)^2}{(1+y^2-2y\cos\phi)^2}
    \;\theta(\tilde{a}>\kappa)
\\&~~
  = \frac{1}{2\pi} \int_{\kappa}^2 d\tilde{a}
    \;2\int_0^{\arccos(\tilde{a}/2)} d\theta
    \left[\tilde{a}-4\cos\theta+\frac{4\cos^2\theta}{\tilde{a}}\right]
\\&~~
  = \frac{1}{\pi} \int_{\kappa}^2 d\tilde{a} \left[\tilde{a}\arccos(\tilde{a}/2)
    - 3\sqrt{1-\tilde{a}^2/4}+\frac{2}{\tilde{a}}\arccos(\tilde{a}/2)\right]
  = \ln\frac{1}{\kappa} - 1 + {\cal O}(\kappa).
\end{split}
\end{equation}
Combining both parts together we obtain:
\begin{equation}
\begin{split}
&G^{CT}_{\tilde{a}>\kappa} =
 \ln\frac{Q}{q_0}\;
 \frac{8C_AC_F}{(1-x)^4} 
 \left(\frac{\alpha_S}{2\pi}\right)^2
 \int\limits_{1-x=\alpha_1+\alpha_2} d\alpha_1d\alpha_2\;
 \alpha_1\alpha_2
 \bigg[
       T^{Vg}_{+} \frac{1}{2}\ln\frac{1}{\kappa}
     + T_3^{Vg}\left(\ln\frac{1}{\kappa}-1\right) 
 \bigg].
\end{split}
\end{equation}
After performing the $\alpha$-integration the final result
for the first counterterm of eq.~(\ref{eq:WVgCT}) reads:
\begin{equation}
\begin{split}
&G^{CT}_{\tilde{a}>\kappa} =
 \ln\frac{Q}{q_0}\; \Peu^{CT}_{\tilde{a}>\kappa}(x),
\\&
%  \Peu^{CT}_{\tilde{a}>\kappa}(x)
% =
%  \frac{2C_F\alpha_S}{\pi}
%  \frac{2C_A\alpha_S}{\pi}
%  \frac{1}{8}
%  \bigg[ \frac{1+x^2}{1-x}
%    \Big( I_0+\ln(1-x)-\frac{11}{12}\Big)
%    4 \ln\frac{1}{\kappa} 
%  + \frac{1}{3}(1-x)
%  - \frac{1}{3}\frac{1+x^2}{1-x} 
% \bigg].
 \Peu^{CT}_{\tilde{a}>\kappa}(x)
=
 \left(\frac{\alpha_S}{2\pi}\right)^2
 \left(\frac{1}{2}C_AC_F\right)
 \bigg[ \frac{1+x^2}{1-x}
   \Big( 4I_0+4\ln(1-x)-\frac{11}{3}\Big)
   4 \ln\frac{1}{\kappa}
\\&~~~~~~~~~~~~
 + \frac{4}{3}(1-x)
 - \frac{4}{3}\frac{1+x^2}{1-x} 
\bigg].
\end{split}
\end{equation}

The integration for the
counterterm of eq.~(\ref{eq:VgK}) representing
the double gluon distribution in the Monte Carlo
proceeds for the same phase space
quite similarly and the final results reads:
% \begin{equation}
% \Peu^{K}_{\tilde{a}>\kappa}(x)
%  =
%   \frac{2C_F \alpha_S}{\pi}
%   \frac{2C_A \alpha_S}{\pi}\;
%   \frac{1}{8}
%   \frac{1+x^2}{1-x} 
%   \bigg(
%            I_0 +\ln (1-x)  -\frac{11}{12}
%   \bigg)
%   4\ln\frac{1}{\kappa},
% \end{equation}
\begin{equation}
\Peu^{K}_{\tilde{a}>\kappa}(x)
 =
 \left(\frac{\alpha_S}{2\pi}\right)^2
 \left(\frac{1}{2}C_AC_F\right)
  \frac{1+x^2}{1-x} 
  \bigg(
           4I_0 +4\ln (1-x)  -\frac{11}{3}
  \bigg)
  4\ln\frac{1}{\kappa},
\end{equation}

The difference between the two counterterms is, of course, colliner-convergent
and it reads
% \begin{equation}
% \Peu^{K-CT}(x)
%  =\Peu^{K}_{\tilde{a}>\kappa}(x)-\Peu^{CT}_{\tilde{a}>\kappa}(x)
% =\frac{2C_F\alpha_S}{\pi}
%  \frac{2C_A\alpha_S}{\pi}
%  \frac{1}{8}
% \bigg[
%  - \frac{1}{3}(1-x)
%  + \frac{1}{3}\frac{1+x^2}{1-x} 
% \bigg].
% \end{equation}
\begin{equation}
\Peu^{K-CT}(x)
 =\Peu^{K}_{\tilde{a}>\kappa}(x)-\Peu^{CT}_{\tilde{a}>\kappa}(x)
=
 \left(\frac{\alpha_S}{2\pi}\right)^2
 \left(\frac{1}{2}C_AC_F\right)
\bigg[
 \frac{4}{3}\frac{1+x^2}{1-x}
 - \frac{4}{3}(1-x)
\bigg].
\end{equation}
The above is used to 
correct the Vg subtracted result of eq.~(\ref{eq:VgSubCT9})
in order to obtain the result of eq.~(\ref{eq:PVgSub}).

% \vspace{5mm}\noindent
% [[[[[[[[[[[[[[[[[[[[[[[[[[[[[[[[[[[[[[[[[[[[[[[[[[[[[[[[[[[[\\
% {\bf TEMPORARY PART, to be deleted?}
% Adding Vg-CT and $G^{CT}_{a>\kappa a_{\max}}$ the non-soft
% un-subtracted contribution of Vg diagram to PDF is obtained:
% \begin{equation}
% \begin{split}
% &G^{Vg}_{a>\kappa a_{\max}} =
%   \ln\frac{Q}{q_0}\;
%   \frac{2C_F \alpha_S}{\pi}
%   \frac{2C_A \alpha_S}{\pi}
% %  \left(\frac{1}{2}C_AC_F\right)
% %  \left(\frac{\alpha_S}{2\pi}\right)^2
%  \frac{1}{8}
%  \bigg\{
%     \frac{1+x^2}{1-x}\bigg[
%        \bigg( I_0+\ln(1-x)-\frac{11}{12}\bigg)
%          4\ln\frac{1}{\kappa}
% \\&~~~
%   - 2I_1 - 2I_0 + 2I_0\ln(1-x)
%   - 2I_0\ln(x) + \ln^2(1-x) - 2\ln(x)\ln(1-x)
% \\&~~~
%   - 2\ln(1-x)
%   + \frac{11}{6}\ln(x) + \frac{\pi^2}{3} - \frac{16}{9}
%     \bigg] 
%   - \frac{1}{6}(1-x) 
% \bigg\}.
% \end{split}
% \end{equation}
% ]]]]]]]]]]]]]]]]]]]]]]]]]]]]]]]]]]]]]]]]]]]]]]]]]]]]]]]]]]]

%
%=================================================
\subsection{2R collinear singularity of Vg in $n$-dimensions}
\label{app:Vg_unres}
For the purpose of combining the 2R contribution with virtual
corrections (gluon self-energy) it is useful
to calculate the contribution to  $\Gamma$, bare PDF,
or PDF of the MC in $n=4+2\epsilon$ dimensions,
in the collinear region $|\ba|<\kappa a_{\max}$.
Let us start from the expression 
where dimensionless variables
$y_i=a_i/a_{\max}$ are already introduced 
and the integration over $a_{\max}$ is performed,
giving the $\frac{1}{\epsilon}$ factor:
\begin{equation}
\begin{split}
&\Gamma^{CT}_{\tilde{a}<\kappa} 
= \text{PP}
\bigg\{
  \frac{4C_AC_F}{(1-x)^2}
  \frac{Q^{4\epsilon}}{\mu^{4\epsilon}}
  \frac{1}{4\epsilon}
  \left(\frac{\alpha_S}{2\pi}\right)^2
  \frac{\Omega_{2+2\epsilon}}{(2\pi)^{2+4\epsilon}}
  \int \frac{d\alpha_1}{\alpha_1}\frac{d\alpha_2}{\alpha_2}
     (\alpha_1\alpha_2)^{2\epsilon}
  \delta_{1-x=\alpha_1+\alpha_2}
\\&~~~\times
     \int d\Omega_{2+2\epsilon} \bigg[
     \int_0^1 dy_1
     \frac{y_1}{\tilde{q}^4(1,1)}
     \bigg( \frac{T_{+}^{Vg}(\epsilon)}{\tilde{a}^2(y_1,1)}
     + (1+\epsilon)T_3^{Vg}\frac{(y_1^2-1)^2}{\tilde{a}^4(y_1,1)} \bigg)
\\&~~~~~~~~~~~~~~~~ 
   + \int_0^1 dy_2
     \frac{y_2}{\tilde{q}^4(1,1)}
     \bigg( \frac{T_{+}^{Vg}(\epsilon)}{\tilde{a}^2(1,y_2)}
     + (1+\epsilon)T_3^{Vg}\frac{(1-y_2^2)^2}{\tilde{a}^4(1,y_2)}
     \bigg) \bigg]\theta_{\tilde{a}<\kappa}
\bigg\}.
\end{split}
\end{equation}
The two integrals over $y_1$ and $y_2$ are equal, hence:
\begin{equation}
\begin{split}
&\Gamma^{CT}_{\tilde{a}<\kappa} 
= \text{PP}
\bigg\{
     \frac{4C_AC_F}{(1-x)^4}\frac{Q^{4\epsilon}}{\mu^{4\epsilon}}
     \frac{1}{4\epsilon}
     \left(\frac{\alpha_S}{2\pi}\right)^2
     \frac{\Omega_{2\epsilon}\Omega_{2+2\epsilon}}{(2\pi)^{2+4\epsilon}}
     \int d\alpha_1d\alpha_2
     (\alpha_1\alpha_2)^{1+2\epsilon}
   \delta_{1-x=\alpha_1+\alpha_2}
\\&~~~~\times
     2\int_0^{\pi} d\theta (\sin\theta)^{2\epsilon}
     \int_0^1 dy \;y
     \bigg( \frac{T_{+}^{Vg}(\epsilon)}{\tilde{a}^2(y,1)}
     + (1+\epsilon)T_3^{Vg}\frac{(y^2-1)^2}{\tilde{a}^4(y,1)} \bigg)
    \theta_{\tilde{a}<\kappa}
\bigg\}.
\end{split}
\end{equation}
One more time we see that the $\alpha$ integration factorizes.
We start the integration with $y$ and $\phi$ and calculate
them separately for the $1/\tilde{a}^2$ and $1/\tilde{a}^4$ parts.
We use the variables $(\tilde{a},\theta)$ introduced before
instead of $(y,\phi)$, then:
\begin{equation}
\begin{split}
&J_1 = 
  2\int\limits_0^{\pi} d\phi 
  \int\limits_0^1 dy\;
  \frac{y(\sin\phi)^{2\epsilon}}{\tilde{a}^2(y,1)}
  \theta_{\tilde{a}<\kappa} 
= 2\int\limits_0^{\kappa} d\tilde{a}
  \int\limits_0^{\arccos(\tilde{a}/2)} d\theta \;
  \tilde{a}^{2\epsilon-1} \left(\frac{\sin\theta}
  {\sqrt{1+\tilde{a}^2-2\tilde{a}\cos\theta}}\right)^{2\epsilon}.
\end{split}
\end{equation}
The $\epsilon$ pole is extracted in form of $\tilde{a}^{2\epsilon-1}$
so we can perform the $\epsilon$ expansion of the rest of
the above expression.
We also use an additional approximation, since $a$ is smaller
then the cutoff $\kappa$ and we are only interested in
logs of $\kappa$ we can fix the integration limits of
the theta integral as $0$ and $\pi/2$, then:
\begin{equation}
J_1 = 2\int_0^{\kappa} d\tilde{a} \;\tilde{a}^{2\epsilon-1}
      \bigg[ \frac{\pi}{2} + \epsilon\bigg( 2\tilde{a}
      - \pi\ln(2) \bigg) \bigg]
    = \pi\left(\frac{1}{2\epsilon} + \ln(\kappa)
    - \ln(2) \right) + {\cal O}(\kappa) + {\cal O}(\epsilon),
\end{equation}
where we performed expansions in $\epsilon$ and $\kappa$.
Performing the same operations for the $1/\tilde{a}^4$ term
(the same order for expansions and integrations) we obtain:
\begin{equation}
\begin{split}
&J_2 = 2\int_0^{\pi} d\phi \int_0^1 dy\;
    \frac{y(1-y^2)^2(\sin\phi)^{2\epsilon}}{\tilde{a}^4(y,1)}
    \theta(\tilde{a}<\kappa)\\
&\;\;\; = 2\int_0^{\kappa} d\tilde{a} \int_0^{\pi/2}
    d\theta \; \tilde{a}^{2\epsilon-1} (2\cos\theta-\tilde{a})^2
    \left(\frac{\sin\theta}{\sqrt{1+\tilde{a}^2-2\tilde{a}\cos\theta}}\right)^{2\epsilon}\\
&\;\;\; = 2\pi\left(\frac{1}{2\epsilon} + \ln(\kappa) - \ln(2)
    - \frac{1}{2}\right) + {\cal O}(\kappa) + {\cal O}(\epsilon).
\end{split}
\end{equation}
Finally,
we sum up both contributions and perform the $\alpha$
integration obtaining:
% \begin{equation}
% \begin{split}
% &\Gamma^{CT}_{\tilde{a}<\kappa} 
%   = \frac{1}{2\epsilon}
%     \frac{2C_F\alpha_S}{\pi}
%     \frac{2C_A\alpha_S}{\pi}
% \\&~~~\times
%  \frac{1}{8}
%  \bigg\{
%       \frac{1+x^2}{1-x}\bigg[ \bigg(\frac{1}{2\epsilon}
%     + \ln\left(\frac{Q^2}{4\pi\mu^2}\right) + \gamma
%     - \ln\frac{1}{\kappa} \bigg)
%     2\Big(I_0+\ln(1-x)-\frac{11}{12}\Big)
% \\&~~~~~~~~~
%     + (1-x)\left(I_0+\ln(1-x)-\frac{11}{12}\right)
% \\&~~~~~~~~~
%     + 2I_1 + 2I_0\ln(1-x) + 3\ln^2(1-x)
%     - \frac{11}{3}\ln(1-x) - \frac{\pi^2}{3}
%     + \frac{67}{18} \bigg]
% \bigg\}.
% \end{split}
% \end{equation}
\begin{equation}
\begin{split}
&\Gamma^{CT}_{\tilde{a}<\kappa} 
  = \frac{1}{2\epsilon}
    \left(\frac{\alpha_S}{2\pi}\right)^2
    \left(\frac{1}{2}C_AC_F\right)
\\&~~~\times
 \bigg\{
      \frac{1+x^2}{1-x}\bigg[ \bigg(\frac{1}{2\epsilon}
    + \ln\left(\frac{Q^2}{4\pi\mu^2}\right) + \gamma
    - \ln\frac{1}{\kappa} \bigg)
    2\Big(4I_0+4\ln(1-x)-\frac{11}{3}\Big)
\\&~~~~~~~~~
    + 8I_1 + 8I_0\ln(1-x) + 12\ln^2(1-x)
    - \frac{44}{3}\ln(1-x) - \frac{4\pi^2}{3}
    + \frac{134}{9} \bigg]
\\&~~~~~~~~~
    + (1-x)\left(4I_0+4\ln(1-x)-\frac{11}{3}\right)
\bigg\}.
\end{split}
\end{equation}
The above result will be useful for combining the 2R
contribution from Vg with the virtual corrections.

% \newpage
%%%%%%%%%%%%%%%%%%%%%%%%%%%%%%%%%%%%%%%%%%%%%%%%%%%%%%%%%%%%%%%%%%%%%%%%%%%%
\bibliographystyle{JHEP}
%%%%\bibliographystyle{h-physrev3}
% \bibliography{radcor}

\begin{thebibliography}{10}

\bibitem{Gross:1973ju}
D.~J. Gross and F.~Wilczek, {\it {Asymptotically Free Gauge Theories. 1}},
  {\em Phys. Rev.} {\bf D8} (1973) 3633--3652.

\bibitem{Gross:1974cs}
D.~J. Gross and F.~Wilczek, {\it {Asymptotically Free Gauge Theories. 2}},
  {\em Phys. Rev.} {\bf D9} (1974) 980--993.

\bibitem{Georgi:1951sr}
H.~Georgi and H.~D. Politzer, {\it {Electroproduction scaling in an
  asymptotically free theory of strong interactions}},  {\em Phys. Rev.} {\bf
  D9} (1974) 416--420.

\bibitem{Ellis:1978ty}
R.~K. Ellis, H.~Georgi, M.~Machacek, H.~D. Politzer, and G.~G. Ross, {\it
  {Perturbation Theory and the Parton Model in QCD}},  {\em Nucl. Phys.} {\bf
  B152} (1979) 285.

\bibitem{Collins:1984kg}
J.~C. Collins, D.~E. Soper, and G.~Sterman, {\it Transverse momentum
  distribution in Drell-Yan pair and W and Z boson production},  {\em Nucl.
  Phys.} {\bf B250} (1985) 199.

\bibitem{Bodwin:1984hc}
G.~T. Bodwin, {\it Factorization of the Drell-Yan cross-section in perturbation
  theory},  {\em Phys. Rev.} {\bf D31} (1985) 2616.

\bibitem{DGLAP}
L.N. Lipatov, {\em Sov. J. Nucl. Phys.} {\bf 20} (1975) 95;\\ V.N. Gribov and
  L.N. Lipatov, {\em Sov. J. Nucl. Phys.} {\bf 15} (1972) 438;\\ G. Altarelli
  and G. Parisi, {\em Nucl. Phys.} {\bf 126} (1977) 298;\\ Yu. L. Dokshitzer,
  {\em Sov. Phys. JETP} {\bf 46} (1977) 64.

\bibitem{Floratos:1978ny}
E.~G. Floratos, D.~A. Ross, and C.~T. Sachrajda, {\it Higher order effects in
  asymptotically free gauge theories. 2. flavor singlet Wilson operators and
  coefficient functions},  {\em Nucl. Phys.} {\bf B152} (1979) 493.

\bibitem{Curci:1980uw}
G.~Curci, W.~Furmanski, and R.~Petronzio, {\it Evolution of parton densities
  beyond leading order: The nonsinglet case},  {\em Nucl. Phys.} {\bf B175}
  (1980) 27.

\bibitem{Vogt:2004mw}
A.~Vogt, S.~Moch, and J.~A.~M. Vermaseren, {\it The three-loop splitting
  functions in QCD: The singlet case},  {\em Nucl. Phys.} {\bf B691} (2004)
  129--181, [\href{http://xxx.lanl.gov/abs/hep-ph/0404111}{{\tt
  hep-ph/0404111}}].

\bibitem{Moch:2004pa}
S.~Moch, J.~A.~M. Vermaseren, and A.~Vogt, {\it The three-loop splitting
  functions in QCD: The non-singlet case},  {\em Nucl. Phys.} {\bf B688} (2004)
  101--134, [\href{http://xxx.lanl.gov/abs/hep-ph/0403192}{{\tt
  hep-ph/0403192}}].

\bibitem{Sjostrand:1985xi}
T.~Sjostrand, {\it A model for initial state parton showers},  {\em Phys.
  Lett.} {\bf B157} (1985) 321.

\bibitem{Webber:1984if}
B.~R. Webber, {\it A QCD model for jet fragmentation including soft gluon
  interference},  {\em Nucl. Phys.} {\bf B238} (1984) 492.

\bibitem{Kulesza:1999sg}
A.~Kulesza and W.~J. Stirling, {\it {On the resummation of subleading
  logarithms in the transverse momentum distribution of vector bosons produced
  at hadron colliders}},  {\em JHEP} {\bf 01} (2000) 016,
  [\href{http://xxx.lanl.gov/abs/hep-ph/9909271}{{\tt hep-ph/9909271}}].

\bibitem{Kulesza:2002rh}
A.~Kulesza, G.~F. Sterman, and W.~Vogelsang, {\it {Joint resummation in
  electroweak boson production}},  {\em Phys. Rev.} {\bf D66} (2002) 014011,
  [\href{http://xxx.lanl.gov/abs/hep-ph/0202251}{{\tt hep-ph/0202251}}].

\bibitem{Marchesini:1988cf}
G.~Marchesini and B.~R. Webber, {\it Monte Carlo simulation of general hard
  processes with coherent QCD radiation},  {\em Nucl. Phys.} {\bf B310} (1988)
  461.

\bibitem{Altarelli:1979ub}
G.~Altarelli, R.~K. Ellis, and G.~Martinelli, {\it {Large Perturbative
  Corrections to the Drell-Yan Process in QCD}},  {\em Nucl. Phys.} {\bf B157}
  (1979) 461.

\bibitem{Anastasiou:2003ds}
C.~Anastasiou, L.~J. Dixon, K.~Melnikov, and F.~Petriello, {\it High-precision
  QCD at hadron colliders: Electroweak gauge boson rapidity distributions at
  NNLO},  {\em Phys. Rev.} {\bf D69} (2004) 094008,
  [\href{http://xxx.lanl.gov/abs/hep-ph/0312266}{{\tt hep-ph/0312266}}].

\bibitem{Frixione:2002ik}
S.~Frixione and B.~R. Webber, {\it Matching NLO QCD computations and parton
  shower simulations},  {\em JHEP} {\bf 06} (2002) 029,
  [\href{http://xxx.lanl.gov/abs/hep-ph/0204244}{{\tt hep-ph/0204244}}].

\bibitem{Jadach:2011cr}
S.~Jadach, A.~Kusina, W.~Placzek, M.~Skrzypek, and M.~Slawinska, {\it {On the
  inclusion of the QCD NLO corrections in the quark-- gluon Monte Carlo
  shower}},  \href{http://xxx.lanl.gov/abs/1103.5015}{{\tt arXiv:1103.5015}}.

\bibitem{Ward:2007xc}
B.~F.~L. Ward, {\it {IR-Improved DGLAP Theory: Kernels, Parton Distributions,
  Reduced Cross Sections}},  {\em Annals Phys.} {\bf 323} (2008) 2147--2171,
  [\href{http://xxx.lanl.gov/abs/0707.3424}{{\tt arXiv:0707.3424}}].

\bibitem{Joseph:2009rh}
S.~Joseph, S.~Majhi, B.~F.~L. Ward, and S.~A. Yost, {\it {HERWIRI1.0: MC
  Realization of IR-Improved DGLAP-CS Parton Showers}},  {\em Phys. Lett.} {\bf
  B685} (2010) 283--292, [\href{http://xxx.lanl.gov/abs/0906.0788}{{\tt
  arXiv:0906.0788}}].

\bibitem{Joseph:2010cq}
S.~Joseph, S.~Majhi, B.~F.~L. Ward, and S.~A. Yost, {\it {New Approach to
  Parton Shower MC's for Precision QCD Theory: HERWIRI1.0(31)}},  {\em Phys.
  Rev.} {\bf D81} (2010) 076008, [\href{http://xxx.lanl.gov/abs/1001.1434}{{\tt
  arXiv:1001.1434}}].

\bibitem{Collins:1998rz}
J.~C. Collins, {\it Hard-scattering factorization with heavy quarks: A general
  treatment},  {\em Phys. Rev.} {\bf D58} (1998) 094002,
  [\href{http://xxx.lanl.gov/abs/hep-ph/9806259}{{\tt hep-ph/9806259}}].

\bibitem{Altarelli:1984pt}
G.~Altarelli, R.~K. Ellis, M.~Greco, and G.~Martinelli, {\it {Vector Boson
  Production at Colliders: A Theoretical Reappraisal}},  {\em Nucl. Phys.} {\bf
  B246} (1984) 12.

\bibitem{Nason:2004rx}
P.~Nason, {\it A new method for combining NLO QCD with shower Monte Carlo
  algorithms},  {\em JHEP} {\bf 11} (2004) 040,
  [\href{http://xxx.lanl.gov/abs/hep-ph/0409146}{{\tt hep-ph/0409146}}].

\bibitem{Collins:2007ph}
J.~C. Collins, T.~C. Rogers, and A.~M. Stasto, {\it {Fully Unintegrated Parton
  Correlation Functions and Factorization in Lowest Order Hard Scattering}},
  {\em Phys. Rev.} {\bf D77} (2008) 085009,
  [\href{http://xxx.lanl.gov/abs/0708.2833}{{\tt arXiv:0708.2833}}].

\bibitem{Jadach:2010ew}
S.~Jadach, M.~Skrzypek, A.~Kusina, and M.~Slawinska, {\it {Exclusive Monte
  Carlo modelling of NLO DGLAP evolution}},  {\em PoS} {\bf RADCOR2009} (2010)
  069, [\href{http://xxx.lanl.gov/abs/1002.0010}{{\tt arXiv:1002.0010}}].

\bibitem{Jadach:2010aa}
S.~Jadach, A.~Kusina, M.~Skrzypek, and M.~Slawinska, {\it {Monte Carlo
  modelling of NLO DGLAP QCD evolution in the fully unintegrated form}},  {\em
  Nucl. Phys. Proc. Suppl.} {\bf 205-206} (2010) 295--300,
  [\href{http://xxx.lanl.gov/abs/1007.2437}{{\tt arXiv:1007.2437}}].

\bibitem{yfs:1961}
D.~R. Yennie, S.~Frautschi, and H.~Suura {\em Ann. Phys. (NY)} {\bf 13} (1961)
  379.

\bibitem{khoze-book}
Y.~Dokshitzer, V.~Khoze, A.~Mueller, and S.~Troyan, {\em Basics of Perturbative
  QCD}.
\newblock Editions Frontieres, 1991.

\bibitem{Slawinska:2009gn}
M.~Slawinska and A.~Kusina, {\it {Non-abelian infra-red cancellations in the
  unintegrated NLO kernel}},  {\em Acta Phys. Polon.} {\bf B40} (2009)
  2097--2108, [\href{http://xxx.lanl.gov/abs/0905.1403}{{\tt
  arXiv:0905.1403}}].

\bibitem{Kusina:2010gp}
A.~Kusina, S.~Jadach, M.~Skrzypek, and M.~Slawinska, {\it {Properties of
  inclusive versus exclusive QCD evolution kernels}},  {\em Acta Phys. Polon.}
  {\bf B41} (2010) 1683--1692, [\href{http://xxx.lanl.gov/abs/1004.4131}{{\tt
  arXiv:1004.4131}}].

\bibitem{Everett-1972}
C.~Everett and E.~Cashwell, {\it Monte Carlo sampler},  1972.
\newblock Los Alamos Report: LA--5061-MS.

\bibitem{Jadach:2005bf}
S.~Jadach and M.~Skrzypek, {\it Solving constrained markovian evolution in QCD
  with the help of the non-markovian Monte Carlo},  {\em Comput. Phys. Commun.}
  {\bf 175} (2006) 511--527,
  [\href{http://xxx.lanl.gov/abs/hep-ph/0504263}{{\tt hep-ph/0504263}}].

\bibitem{Jadach:2005yq}
S.~Jadach and M.~Skrzypek, {\it Non-markovian Monte Carlo algorithm for the
  constrained markovian evolution in QCD},  {\em Acta Phys. Polon.} {\bf B36}
  (2005) 2979--3022, [\href{http://xxx.lanl.gov/abs/hep-ph/0504205}{{\tt
  hep-ph/0504205}}].

\bibitem{Kusina:2011xh}
A.~Kusina, S.~Jadach, M.~Skrzypek, and M.~Slawinska, {\it {NLO evolution
  kernels: Monte Carlo versus MSbar}},
  {\em Acta Phys. Polon.}  {\bf B42} (2011) 1475,
  \href{http://xxx.lanl.gov/abs/1106.1787}{{\tt arXiv:1106.1787}}.

\bibitem{Ellis:1978sf}
R.~K. Ellis, H.~Georgi, M.~Machacek, H.~D. Politzer, and G.~G. Ross, {\it
  Factorization and the parton model in QCD},  {\em Phys. Lett.} {\bf B78}
  (1978) 281.

\bibitem{GolecBiernat:2006xw}
K.~Golec-Biernat, S.~Jadach, W.~P\l{}aczek, and M.~Skrzypek, {\it {Markovian
  Monte Carlo solutions of the NLO QCD evolution equations}},  {\em Acta Phys.
  Polon.} {\bf B37} (2006) 1785--1832,
  [\href{http://xxx.lanl.gov/abs/hep-ph/0603031}{{\tt hep-ph/0603031}}].

\bibitem{Jadach:2007qa}
S.~Jadach, W.~P\l{}aczek, M.~Skrzypek, P.~Stephens, and Z.~Was, {\it
  Constrained MC for QCD evolution with rapidity ordering and minimum k(t)},
  \href{http://xxx.lanl.gov/abs/hep-ph/0703281}{{\tt hep-ph/0703281}}.

\bibitem{Catani:1996vz}
S.~Catani and M.~H. Seymour, {\it {A general algorithm for calculating jet
  cross sections in NLO QCD}},  {\em Nucl. Phys.} {\bf B485} (1997) 291--419,
  [\href{http://xxx.lanl.gov/abs/hep-ph/9605323}{{\tt hep-ph/9605323}}].

\bibitem{GehrmannDeRidder:2005cm}
A.~Gehrmann-De~Ridder, T.~Gehrmann, and E.~W.~N. Glover, {\it {Antenna
  Subtraction at NNLO}},  {\em JHEP} {\bf 09} (2005) 056,
  [\href{http://xxx.lanl.gov/abs/hep-ph/0505111}{{\tt hep-ph/0505111}}].

\bibitem{Heinrich:1997kv}
G.~Heinrich and Z.~Kunszt, {\it {Two-loop anomalous dimension in light-cone
  gauge with Mandelstam-Leibbrandt prescription}},  {\em Nucl. Phys.} {\bf
  B519} (1998) 405--432, [\href{http://xxx.lanl.gov/abs/hep-ph/9708334}{{\tt
  hep-ph/9708334}}].

\bibitem{Bassetto:1998uv}
A.~Bassetto, G.~Heinrich, Z.~Kunszt, and W.~Vogelsang, {\it {The light-cone
  gauge and the calculation of the two-loop splitting functions}},  {\em Phys.
  Rev.} {\bf D58} (1998) 094020,
  [\href{http://xxx.lanl.gov/abs/hep-ph/9805283}{{\tt hep-ph/9805283}}].

\bibitem{foam:2002}
S.~Jadach, {\it Foam: A general purpose cellular Monte Carlo event generator},
  {\em Comput. Phys. Commun.} {\bf 152} (2003) 55--100,
  [\href{http://xxx.lanl.gov/abs/physics/0203033}{{\tt physics/0203033}}].

\bibitem{Slawinska:2010jn}
M.~Slawinska and S.~Jadach, {\it {MCdevelop - the universal framework for
  Stochastic Simulations}},  {\em Comput. Phys. Commun.} {\bf 182} (2011)
  748--762, [\href{http://xxx.lanl.gov/abs/1006.5633}{{\tt arXiv:1006.5633}}].

\end{thebibliography}
%%%%%%%%%%%%%%%%%%%%%%%%%%%%%%%%%%%%%%%%%%%%%%%%%%%%%%%%%%%%%%%%%%%%%%%%%%%

\providecommand{\href}[2]{#2}\begingroup\raggedright\endgroup

%%%%%%%%%%%%%%%%%%%%%%%%%%%%%%%
\end{document}